\pdfoutput=1
\documentclass[graybox, envcountchap]{Template/svmult}

\newif\ifarxiv
\arxivtrue
\ifarxiv\usepackage{natbib}\else\usepackage[numbers,sort&compress]{natbib}\fi
\newcommand{\pcite}[1]{\ifarxiv\citealp{#1}\else\citep{#1}\fi}%
\newcommand{\chapterorreview}{\ifarxiv review\else chapter\fi}%
\ifarxiv
\makeatletter
\renewcommand\section{\@startsection{section}{1}{\z@}%
                       {-26\p@}%
                       {12\p@}%
                       {\normalfont\secsize\secstyle
                        \rightskip=\z@ \@plus 8em\pretolerance=10000 }}
\renewcommand\subsection{\@startsection{subsection}{2}{\z@}%
                       {-20\p@}%
                       {11\p@}%
                       {\normalfont\subsecsize\subsecstyle
                        \rightskip=\z@ \@plus 8em\pretolerance=10000 }}
\renewcommand\subsubsection{\@startsection{subsubsection}{3}{\z@}%
                       {-16\p@}%
                       {6\p@}%
                       {\normalfont\normalsize\subsubsecstyle
                        \rightskip=\z@ \@plus 8em\pretolerance=10000 }}
\makeatother
\fi
\ifarxiv
\newenvironment{keeptogether}{\par}{\par}
\else

\fi
\newcommand{\printreferences}{\clearpage\bibliographystyle{aasjournal}\bibliography{ref}}
\usepackage{pifont}
\usepackage{tikz}
\usetikzlibrary{shapes.geometric, arrows.meta, positioning}
\usepackage{mathptmx}
\usepackage{amsmath}
\usepackage{amssymb}
\usepackage{color}
\usepackage{colortbl}
\usepackage{rotating}
\definecolor{rowhl}{rgb}{0.95,0.92,0.84}
\usepackage{helvet}
\usepackage{courier}

\usepackage{makeidx}
\usepackage{graphicx}
\usepackage{subfig}
\usepackage{array}

\usepackage{multicol}
\usepackage[bottom]{footmisc}%

\usepackage{hyperref}
\hypersetup{colorlinks=true, linkcolor=blue, citecolor=blue, urlcolor=blue}
\usepackage{marvosym}
\usepackage{orcidlink}
\providecommand{\fnm}[1]{#1}\providecommand{\sur}[1]{#1}%
\makeindex

\usepackage{longtable}%

\renewenvironment{svgraybox}{\par\addvspace{10pt}\noindent\begin{minipage}{\linewidth}\fboxsep=12pt\relax\begin{shaded}\list{}{\leftmargin=6pt\rightmargin=6pt\topsep=0pt\relax}\expandafter\item\parindent=\svparindent\hskip-\listparindent}{\endlist\end{shaded}\end{minipage}\par\addvspace{10pt}}

\begin{document}

\ifarxiv
\title{Graph Neural Networks for Astrophysics and Cosmology}
\else
\title{Graph Neural Networks}
\fi
\titlerunning{Graph Neural Networks for Astrophysics and Cosmology}
\author{\fnm{Christian Kragh} \sur{Jespersen}\,\orcidlink{0000-0002-8896-6496}}
\tocauthor{Christian Kragh Jespersen}%
\authorrunning{C. K. Jespersen}
\institute{\fnm{Christian Kragh} \sur{Jespersen} (\Letter,  \email{christian.jespersen@utoronto.ca}) \at David A. Dunlap Department of Astronomy \& Astrophysics, University of Toronto, Toronto, ON M5S 3H4, Canada \\ Department of Statistical Sciences, University of Toronto, Toronto, ON M5S 3G3, Canada \\ Canadian Institute for Theoretical Astrophysics, University of Toronto, 60 St George Street, Toronto, M5S 3H8, Ontario, Canada \\  Department of Astrophysical Sciences, Princeton University, Princeton, NJ 08544, USA}
\ifarxiv
\makeatletter\@numarttrue\makeatother%
\setcounter{chapter}{1}%
\else
\setcounter{chapter}{3}%
\fi
\maketitle
\ifarxiv
\makeatletter%
\def\thesection{\@arabic\c@section}
\def\thefigure{\@arabic\c@figure}
\def\thetable{\@arabic\c@table}
\def\theequation{\arabic{equation}}
\makeatother
\fi

\abstract{Physics seldom makes breakthroughs when we consider objects only in isolation. Instead we model and measure how objects sit among their neighbours, spatial, temporal, or otherwise: a galaxy in its environment, a halo in its history, a photon hit in a detector. Graphs are ideally suited for encoding exactly these relations, and Graph Neural Networks (GNNs) learn the geometric functions that respect them. This \chapterorreview{} presents the use of GNNs for astrophysics and cosmology, and argues that GNNs often suit physics and physical data better than any alternative: they fit the native geometry of the domain and the signal, enable the incorporation of the symmetries and prior knowledge we already trust and have toiled to build, and stay flexible enough to capture what we do not yet understand. GNNs' physically intuitive workings also render them more interpretable than other deep-learning architectures. We show how we can begin from general principles of the geometry of deep learning and build the graph and the graph neural network from the ground up, and how other architectures, like the convolutional network and the transformer, can be seen as special cases of one unified geometric blueprint. We survey applications across cosmology, galaxy formation, dynamics, and instrumentation; and we chart the frontier. From this survey, we find that only one in six articles outside instrumentation applies GNNs to data rather than just simulations. We also distil a set of best practices, including when \emph{not} to reach for a graph. Throughout, we will prioritise physical intuition and foundational principles over mathematical machinery whenever possible.}

\section{Introduction}
\label{sec:intro}

The history of physics is a story of geometry. From the ellipses of Kepler, through the invariances of Galileo, to the curved spacetime of Einstein, physics often advances less by the collection of facts than by the discovery of the relations or symmetries that render those facts inevitable \citep{Noether1918_invariante_variationsprobleme}. A theory also often earns our trust when it is stated in a form that no choice of coordinates, no relabelling, and no rotation can disturb. Learning algorithms for the physical world must therefore be grounded in geometry.\footnote{Graph theory itself was born of a geometric puzzle: in 1736 Euler reduced the seven bridges of K\"onigsberg to nodes and edges and proved that no stroll crosses each bridge exactly once. Three centuries on, the same abstraction describes the largest structures in the Universe.}

The natural sciences furthermore run, at heart, on relations. We build theories not of the isolated object but of the relation between objects, and the \textbf{graph} holds exactly this structure: objects become nodes, relations become edges. Such data permeate the natural sciences but especially astrophysics. The cosmic web threads galaxies into filaments, the kinematics and positions of Milky Way stars trace its evolution, and neutrino observatories scatter their sensors through large volumes of ice or water. None of these live on regular Euclidean grids, and to force one upon them invents a structure neither the data nor the physics they measure ever had.

Graph Neural Networks (GNNs) are fundamentally \emph{geometric}\footnote{So are \emph{all} DL architectures, but GNNs are more expressly geometric and are typically formulated as being a foundational part of the modern family of geometric deep learning.} in nature, and they harness relational \textbf{inductive biases} for functions on arbitrary domains \citep{Bronstein2017_geometricDL_review, Bronstein2021_geometricDL_book}. They propagate information along the edges through parameterised \textbf{message passing}, and can read out node-, edge-, and graph-level quantities. These ideas already carry large-scale production systems. AlphaFold reasons over chemical residue graphs \citep{Jumper2021_alphafold_protein}, GraphCast forecasts weather better than other leading numerical models \citep{Lam2023_graphcast_weather}, and GNNs serve the travel-time estimates of Google Maps \citep{DerrowPinion2021_eta_googlemaps}. The idea long predates more recent successes and enthusiasm: recursive and recurrent precursors reach back to the late 1990s \citep[see][and references therein]{Scarselli2009_GNN_model}. The message passing view that organises this \chapterorreview{} can also be used to define the convolutional neural network and the transformer as special cases of message passing on specific types of graphs \citep{Velickovic2022_messagepassing_up}. While these architectures are thus more restrictive, they are more efficient on their specialised domains than the generalist GNN.
\begin{svgraybox}
\textbf{Inductive bias.} The set of assumptions a learning algorithm carries before it meets data which then rank one solution above another when both fit the training set equally well. Weight sharing across pixels encodes translation symmetry in the convolutional neural network (CNN). Message passing along local neighbourhood edges encodes permutation symmetry and locality in the GNN. Distinct from inductive \emph{learning}, a mode of deployment defined in \S\ref{subsec:tasks}.
\end{svgraybox}

Excellent reviews already cover deep learning across astronomy and the physical sciences at large \citep{HuertasCompany2023_dawes_review, Ferguson2025_AI_physicalsciences, Ting2026_deeplearning_review}, whereas other reviews cover GNNs in related fields like physics \citep{Shlomi2021_particlephysics_review}. This \chapterorreview{}\footnote{As well as others in the same series as this one.} covers the more specialised case of the use of GNNs in astrophysics and cosmology. We will take on a geometry-first treatment which will inevitably lead us to graph-specific applications. This spans from a generalised blueprint for considering geometry in learning algorithms from which the common architectures follow, all the way to the practical advice that makes them work on astrophysical data.

This \chapterorreview{} proceeds in three movements: the geometry of learning and the geometric blueprint that unifies the deep learning architectures and motivates the use of graph neural networks (\S\ref{sec:geometry}); the construction of the graph and the graph neural network (\S\ref{sec:graphs}--\ref{sec:gnn}); and then we turn to the more practical side of GNNs in astrophysics: the applications (\S\ref{sec:applications}), the best practices (\S\ref{sec:bestpractices}), and the frontier (\S\ref{sec:frontier}).

\section{Geometry and Deep Learning}
\label{sec:geometry}

\begin{quotation}
\noindent\textit{``The knowledge of certain principles readily supplies the knowledge of certain facts.''} \\
\noindent\hfill --- Claude Adrien Helv\'etius
\end{quotation}

When we master a few foundational principles, we can always \emph{derive}, rather than memorise, the right architecture\footnote{The right architecture will often turn out to be a GNN.} for the problem before us. The recipe takes only a few inputs --- the problem we pose, the signal we aim to measure, the domains on which the data and signal live, and the symmetries to respect --- and then returns the architecture to use. In the following sections, we show how to define the geometry of the data domain, and that of the signal, introduce the necessary language of symmetry, and assemble the blueprint from which any modern deep-learning architecture, be it the CNN, the transformer, or the GNN, can be derived.

These common architectures follow from well-known and general symmetries such as translations, rotations, reflections, or permutations, but many other symmetry groups which we will not cover here --- like the Lorentz group of special relativity --- are astrophysically relevant.
\ifarxiv\enlargethispage{1.5\baselineskip}\else\enlargethispage{2\baselineskip}\fi
\begin{svgraybox}
\textbf{Symmetry group.} A group $G$ is the set of transformations that leave a system recognisably itself. The set is a group if any two transformations composed give a third which is also part of the group, and every transformation has an inverse.\footnotemark{} A capital letter names the kind of transformations, and the number in parentheses typically indicates the dimensions of the space it acts on.
\end{svgraybox}
\footnotetext{Another formal requirement is that ``do nothing'' belongs as the identity. While this is necessary to form a group, it is not very illuminating. \hyperref[app:symmetries]{Appendix~1} supplies more comprehensive formal definitions and the representation theory around them.}

We will often encounter the following classes of groups:

\begin{itemize}
  \item \textbf{T} -- translations: $T(3)$ shifts three-dimensional space.
  \item \textbf{O} -- orthogonal: rotations and reflections that preserve lengths and angles between objects.
  \item \textbf{U} -- unitary: rotations of complex space. It changes phases but not amplitudes.
  \item \textbf{E} -- Euclidean: translations, rotations, and reflections all together.
  \item \textbf{D} -- dihedral: the discrete rotations and flips of a regular
  polygon. The number typically counts the sides of the polygon.
  \item $\Sigma_n$ -- permutations of $n$ objects: the set without a given order.
  \item \textbf{S} -- \emph{special}, a prefix: keep only the orientation-preserving half of a group, with its reflections dropped.
\end{itemize}
A few astrophysically representative groups serve as specific examples:
\begin{itemize}
  \item $SO(3)$ -- rotations of three-dimensional space: isotropy, and full-sky maps on the celestial sphere.
  \item $E(3)$ and $SE(3)$ -- rigid motions of three-dimensional space, with and without reflections: used often by the equivariant architectures of \S\ref{subsec:equivariant-mp}.
\end{itemize}

\subsection{The geometries of the domain, the signal, and the model}
\label{subsec:domain-signal}

\begin{figure}[p]
\centering
\hspace{-0.5cm}
\makebox[\linewidth][c]{\includegraphics[trim={0cm 4.1cm 0cm 1cm},clip, width=1.45\linewidth]{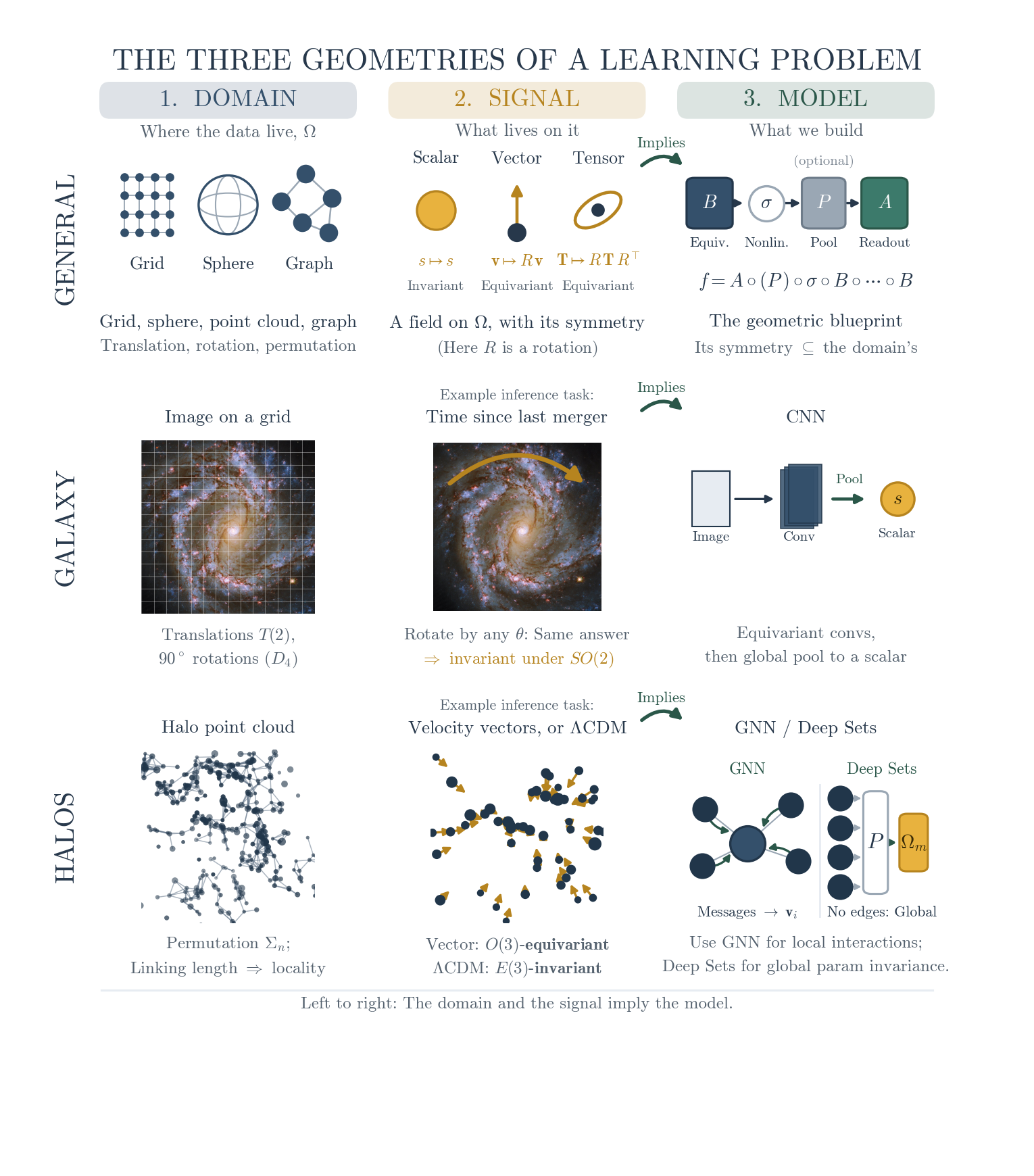}}
\caption{\textbf{The three geometries of a learning problem}. Each column adds one: the \emph{domain} where the data live, the \emph{signal} that exists upon it, and that of the \emph{model}. In a well-motivated case, the domain and signal \emph{imply} the model to be used. The rows fall from the abstract to two concrete cases. A galaxy sits on a pixel grid, whose translations and right-angle rotations a convolutional network inherits before a global pooling returns one rotation-invariant number, here the time since its last merger (but generalisable to most scalars). A halo catalogue forms an irregular point cloud whose only native symmetry is permutation; a linking length adds edges, and with them a bias towards local influences. We should choose the model whose symmetry respects the domain without excessive preprocessing, and is able to fully express the signal. At a minimum, it must not outlaw the true solution. At best, it matches it exactly.}
\label{fig:domain_signal}
\end{figure}

Any learning problem contains three geometries: that of the domain on which the data exist, that of the signal we wish to infer, and that of the model. These are shown in Figure~\ref{fig:domain_signal}. A good model respects the geometries of both the domain and the signal. The geometry of the \emph{domain} comes first: the support on which the data live, a regular grid, a sphere, an irregular point cloud, a graph. The geometry of the \emph{signal} comes second: the structure of the field defined on that support, which carries its own symmetries, a scalar temperature, a vector velocity, a tensor shear, on whatever space we happen to observe it.

The lesson repeats throughout this \chapterorreview{}: choose the algorithm that respects the native geometry of both the domain and the signal (see Table~\ref{tab:gdl_architectures}). Impose no symmetry the data lack, and discard none they do hold. A convolutional network presumes a regular grid. If we run it on e.g., an irregular detector, we must first voxelise the data, thereby inventing a translational structure that does not exist, biasing the learning. Where the domain becomes irregular, the graph is often not a stylistic preference but the only faithful description.

\subsection{Symmetries, invariances, and equivariances}
\label{subsec:symmetries}

\begin{figure}[t]
    \centering
    \includegraphics[trim={0cm 1.7cm 0cm 0cm}, width=0.99\linewidth]{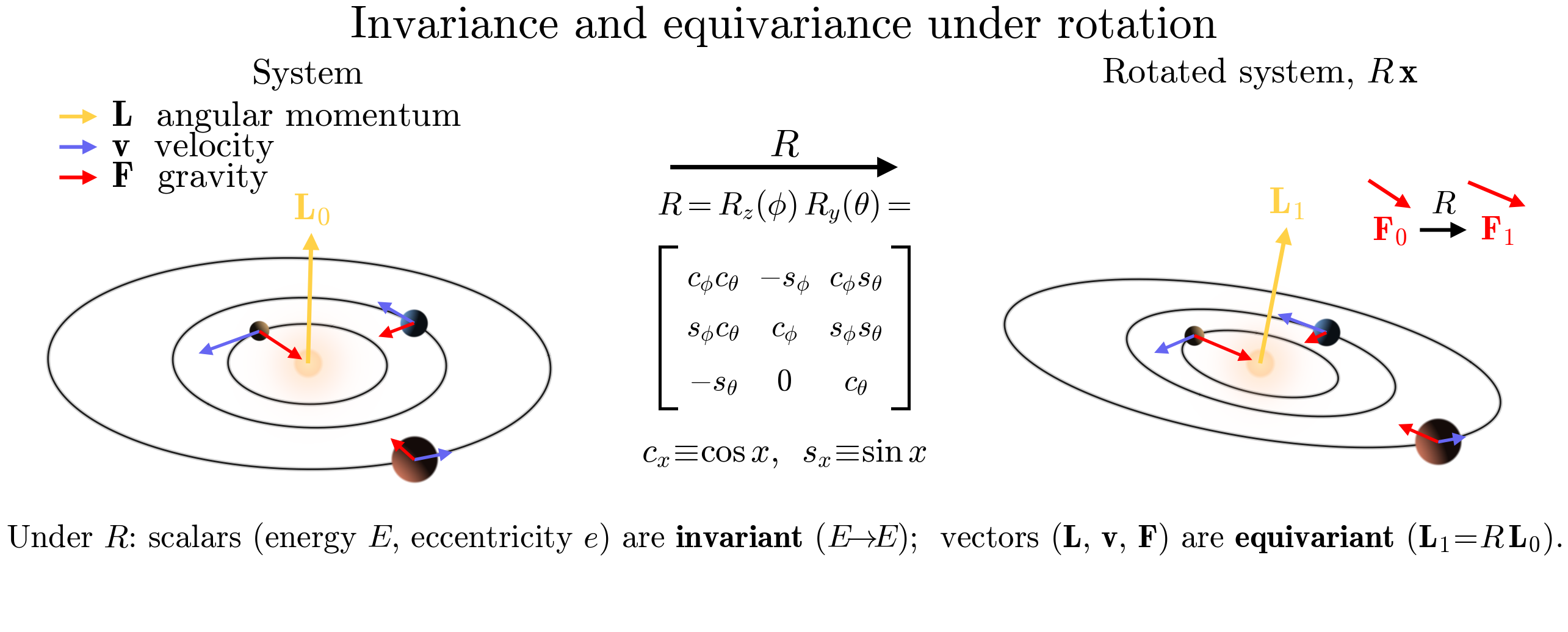}
    \caption{\textbf{Invariances and equivariances in a familiar physical system}. For a planetary system, vector quantities such as forces and velocities are \emph{equivariant} under rotations, meaning that they transform \emph{with} the rotation. Scalar quantities, such as energy or eccentricities, are \emph{invariant} under such transformations. Both respond to the same symmetry, but the distinction must always be kept in mind.}
    \label{fig:rotation_demonstration}
\end{figure}

A \emph{symmetry} occurs when a system or function remains structured and predictable under a specific group of transformations. The physicist meets symmetry early \citep{Weyl1952_symmetry}. The outcome of an experiment may not depend on where the experimental setup stands, which way it faces, or at what time one runs it --- or it may follow these changes exactly. The transformations that define this structure form a \emph{group}, and the way that group acts on our data, through maps we call \emph{representations}, dictates how a sensible model behaves. We keep the detailed formal account, the full group axioms, and the representation theory, to \hyperref[app:symmetries]{Appendix~1}; here two definitions suffice.

\begin{svgraybox}
A function stays \emph{invariant} when the group action leaves its output unchanged, and turns \emph{equivariant} when the output transforms in step with the input. With $g.x$ denoting the action of a transformation $g$ from a group $G$ on an input $x$, and a function $f$, we define:
\begin{itemize}
    \item \textbf{invariance} of $f$ with respect to $G$ when $f(g.x)=f(x)$, for all $g$ and
    \item \textbf{equivariance} of $f$ with respect to $G$ when $f(g.x)=g.f(x)$, for all $g$
\end{itemize}
\end{svgraybox}

\begin{figure}
    \centering
    \includegraphics[width=0.99\linewidth]{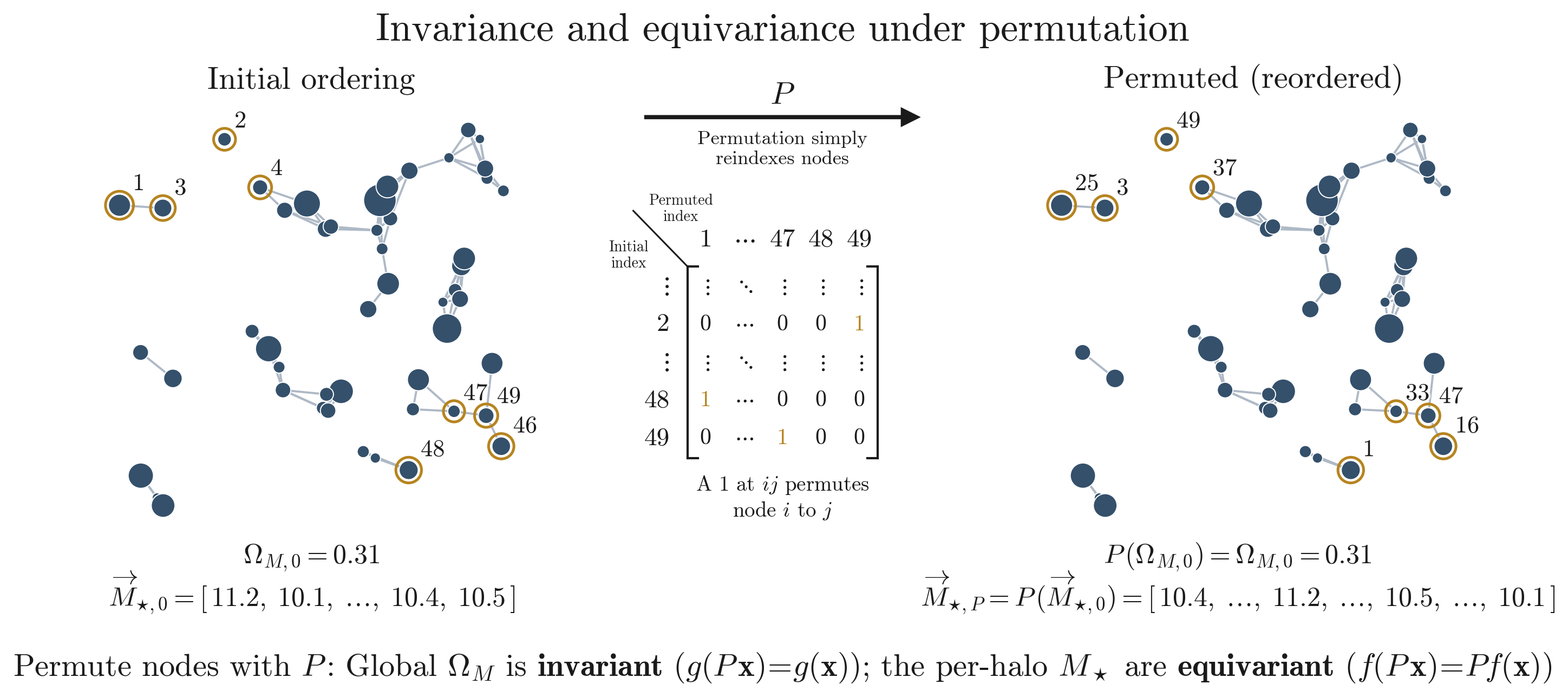}
    \caption{\textbf{Permutation symmetry among objects without \textit{a priori} ordering}. A single cosmic graph --- halos joined by proximity, sized by mass --- carries the same physics under relabelling, with the permutation $P$ that relates the two orderings written as a condensed permutation matrix. Reorder the nodes and the graph holds, but different outputs require either invariance or equivariance: a graph-level read-out such as the cosmic matter density $\Omega_M$ should stay put (invariant), while the per-halo stellar masses $M_\star$ travel with the relabelling (equivariant). A graph neural network must honour both, so that each halo keeps its prediction regardless of its index.}
    \label{fig:permutation_demonstration}
\end{figure}

The distinction also immediately draws a line between scalars and vectors \citep{Villar2021_scalars_universal}. Rotate a molecule: its total energy (a scalar) does not move (invariant), yet the force on each atom (a vector) rotates with it (equivariant), as Figure~\ref{fig:rotation_demonstration} shows for a planetary system. Permutation obeys the same split: relabel a set of identical particles, and a total energy stays put while the list of per-particle forces reorders with the permutation (Figure~\ref{fig:permutation_demonstration}). The same split organises every prediction we make on a graph, and we return to it in \S\ref{sec:gnn}.\footnote{Note that invariance amounts to the special case of equivariance in which the output carries the trivial representation.} The distinction between invariances and equivariances is fundamental --- yet it is often muddled. This confusion can lead our models to either destroy critical relationships or fail to simplify data, leading to poor generalisation and incorrect predictions.

\subsection{The geometric blueprint, and the architectures it implies}
\label{subsec:blueprint}

Every successful modern deep learning architecture amounts to one recipe, applied to a different domain and a different symmetry. Fix a domain $\Omega$ with symmetry group $G$ and assemble four kinds of blocks, after the blueprint of \citet{Bronstein2021_geometricDL_book}. The box below uses a compact notation that the rest of the chapter leans on. The domain $\Omega$ is the set the data live on, with the pixel grid, the sphere, and the graph as our recurring examples. The group $G$ denotes the symmetry group, and $g$ its individual actions, as when a specific rotation turns an image or a permutation reorders a catalogue. Following the definition in \S\ref{subsec:symmetries}, a layer $B$ is equivariant with respect to $G$ when it commutes with every such action, $B(g.x)=g.B(x)$: we can either transform first and then apply the layer or apply the layer first and then transform, the end results agree.

\begin{svgraybox}
\textbf{The Geometric Deep Learning blueprint.} From a domain $\Omega$ with symmetry group $G$ from which the group actions $g$ can act on inputs $x$ as $g.x$, and a signal on it whose own symmetry decides the final output, build:
\begin{itemize}
  \item a \textbf{$G$-equivariant layer} $B$, $B(g.x)=g.B(x)$;
  \item a pointwise \textbf{non-linearity} $\sigma$;
  \item an optional \textbf{local pooling (coarsening)} $P$, which shrinks the domain while it preserves the symmetry;
  \item a \textbf{global pooling} $A$, which reads out a $G$-\emph{invariant} answer, $A(g.x)=A(x)$.
\end{itemize}
Stack the equivariant blocks (with non-linearities and optional coarsening) and, when the signal calls for an invariant answer, finish with the invariant readout (when the signal itself transforms, stop at an equivariant layer instead):
\[
  f \;=\; A \circ \sigma_J \circ B_J \circ P_{J-1}\circ\cdots\circ
          P_1 \circ \sigma_1 \circ B_1 .
\]
\end{svgraybox}

To see the recipe at work, we can derive different neural architectures from first principles. Since we often hold a firm intuition for the image and a shakier one for the graph, let us lead with the more familiar case of the convolutional neural network. We demand three things \emph{a priori} of a model that acts on an image: translational \emph{equivariance} of its internal operations, scale separation, and translational \emph{invariance} of its output. The first demand decides everything. Imagine translating a dog across an image --- the filter which identifies it must clearly follow the dog. To insist that a local linear operation commute with every translation, that is, to insist on translational equivariance, forces it to share weights across the image, that is, to become a \emph{convolution} (\hyperref[app:conv-proof]{Appendix~2} contains the proof). Scale separation, the premise that one may coarsen the signal, justifies the interleaved \emph{pooling} that stacks these convolutions into a multiscale hierarchy; a dog stays a dog across scale. We then finally demand that the \emph{global output} (i.e. is there a dog here or not?) ignore position, and a single \emph{global pooling} delivers it. Reassembled, the parts make exactly the familiar convolutional network, a direct consequence of the blueprint above with the group set to translation.

The same demands, made on a graph, will hand us the GNN. Demand permutation equivariance of the internal operations and locality along the edges: with the group set to $G=\Sigma_n$, the question becomes which layers $B$ commute with every relabelling; the permutation-equivariant linear layer on a set is forced into a form that can be described as ``itself plus one shared aggregation over the rest''. This result is proven as Lemma 3 of the Deep Sets approach of \citet{Zaheer2017_deepsets}, and we redo a version of it in \hyperref[app:conv-proof]{Appendix~2}. Once locality is added through the graph neighbourhood, the result is message passing: the lemma's ``itself'' becomes the node update $\phi$, and the shared aggregation becomes learned messages $\psi$ collected over the neighbourhood by any permutation-invariant aggregator, usually written as $\bigoplus$. This is exactly the layer which \S\ref{sec:gnn} builds in full. If we are doing a task where the signal is a graph-level scalar, we may finally demand a permutation-invariant output, and a global pooling, $A$, delivers it. The CNN and the GNN thus fall out of one recipe, with only the group exchanged. The same recipe, with a different group and a different notion of locality, yields the other model architectures. The transformer becomes the blueprint on a fully connected graph, where the \textbf{attention}\footnote{Attention is the inferred scalar coupling between two objects --- like two words in a paragraph.} coefficients \citep{Vaswani2017_transformer_attention} act as a learned, soft adjacency, and the deep set becomes the blueprint on a set, the edgeless version of a graph. Each of these is more highly optimised for its specialised use case than the general GNN, but they all descend from one unified geometric principle. Table~\ref{tab:gdl_architectures} collects the common architectures as instances of the geometric blueprint.\footnote{Geometric deep learning typically contains five families of domains, the ``five Gs'': the grid, the group, the graph, the geodesic, and the gauge \citep{Bronstein2021_geometricDL_book}. For topological and higher-order domains, see \citet{Papillon2025_beyondeuclid_geometric}.}
\begin{table}[t]
\caption{Common geometric architectures as instances of the geometric deep learning blueprint, after \citet{Bronstein2021_geometricDL_book} (trimmed). The highlighted row marks the GNN, the focus of this \chapterorreview{}.}
\label{tab:gdl_architectures}
\centering\footnotesize\setlength{\tabcolsep}{4pt}\renewcommand{\arraystretch}{1.5}
\makebox[\linewidth][c]{%
\begin{tabular}{llll>{\raggedright\arraybackslash}p{2.6cm}>{\raggedright\arraybackslash}p{2.8cm}}
\hline\noalign{\smallskip}
Architecture & Domain $\Omega$ & Symmetry $G$ & Locality & Example (non-astro) & Example (astronomy) \\
\noalign{\smallskip}\svhline\noalign{\smallskip}
CNN           & Grid           & Translation            & Local patch & Image classification \citep{Krizhevsky2012_alexnet_imagenet} & Galaxy morphology \citep{Dieleman2015_galaxymorphology_CNN} \\
Spherical CNN & Sphere         & Rotation $SO(3)$       & Local cap   & Omnidirectional vision \citep{Cohen2018_spherical_CNN} & Weak-lensing maps \citep{Perraudin2019_deepsphere_spherical} \\
\rowcolor{rowhl}
GNN           & Graph          & Permutation $\Sigma_n$ & Neighbours  & Molecular properties \citep{Gilmer2017_MPNN_quantumchemistry} & Merger trees \citep{Jespersen2022_mangrove} \\
Deep Sets     & Set            & Permutation $\Sigma_n$ & None        & Point clouds \citep{Zaheer2017_deepsets} & Thermal SZ field \citep{Thiele2022_tSZ_deepsets} \\
Transformer   & Complete graph & Permutation $\Sigma_n$ & All-to-all  & Language \citep{Vaswani2017_transformer_attention} & Light curves \citep{DonosoOliva2023_astromer_transformer} \\
\noalign{\smallskip}\hline
\end{tabular}%
}
\end{table}

\section{Graphs}
\label{sec:graphs}

Before we learn on a graph, we will define what a graph is and how to build one. While we focus on graphs for machine learning, the graph picture is valuable even when no neural network touches it. A graph path-likelihood, for one, extracts galaxy-formation information from the structure of the layered halo graph alone, with not a single learned weight in sight \citep{Yang2026_halograph_likelihood_not_gnn}.

\subsection{Definitions and notation}
\label{subsec:graph-defs}

We define the graph as the triplet $G=(V,E,U)$: $V=\{\mathbf{x}_i\}_{i=1:N^v}$ holds the node feature vectors\footnote{The node feature vector is often written $\mathbf{v}_i$ in the machine-learning literature, with $v$ for vertex. We keep $\mathbf{x}_i$ to avoid confusion with velocities.}; $E=\{(\mathbf{e}_k,r_k,s_k)\}$ holds the edge feature vectors, each tied to a receiver $r_k$ and a sender $s_k$; and $U$ holds the global, graph-level feature vector, after the notation of \citet{Battaglia2018_relational_GN}. The connectivity also admits a complementary description through the binary adjacency matrix $A$, with $A_{ij}=1$ to mark the existence of an edge.\footnote{The literature also often writes $u$ for the receiving node and $v$ for a sending neighbour. Here we prefer the physics-native $ij$-indexing.} $A$ is thus not independent from the typical description of $G$, as it can be constructed fully from $E$ without regard to the actual edge features, but the matrix description of $A$ is highly useful for understanding and computation alike. The real graph often runs sparse, so in practice one can obtain a compact description of $A$ by storing not the dense $N^v\times N^v$ matrix but the \emph{edge index}, the two rows of sender and receiver indices.

\begin{figure}[h]
\centering
\makebox[\linewidth][c]{\includegraphics[width=1.2\linewidth]{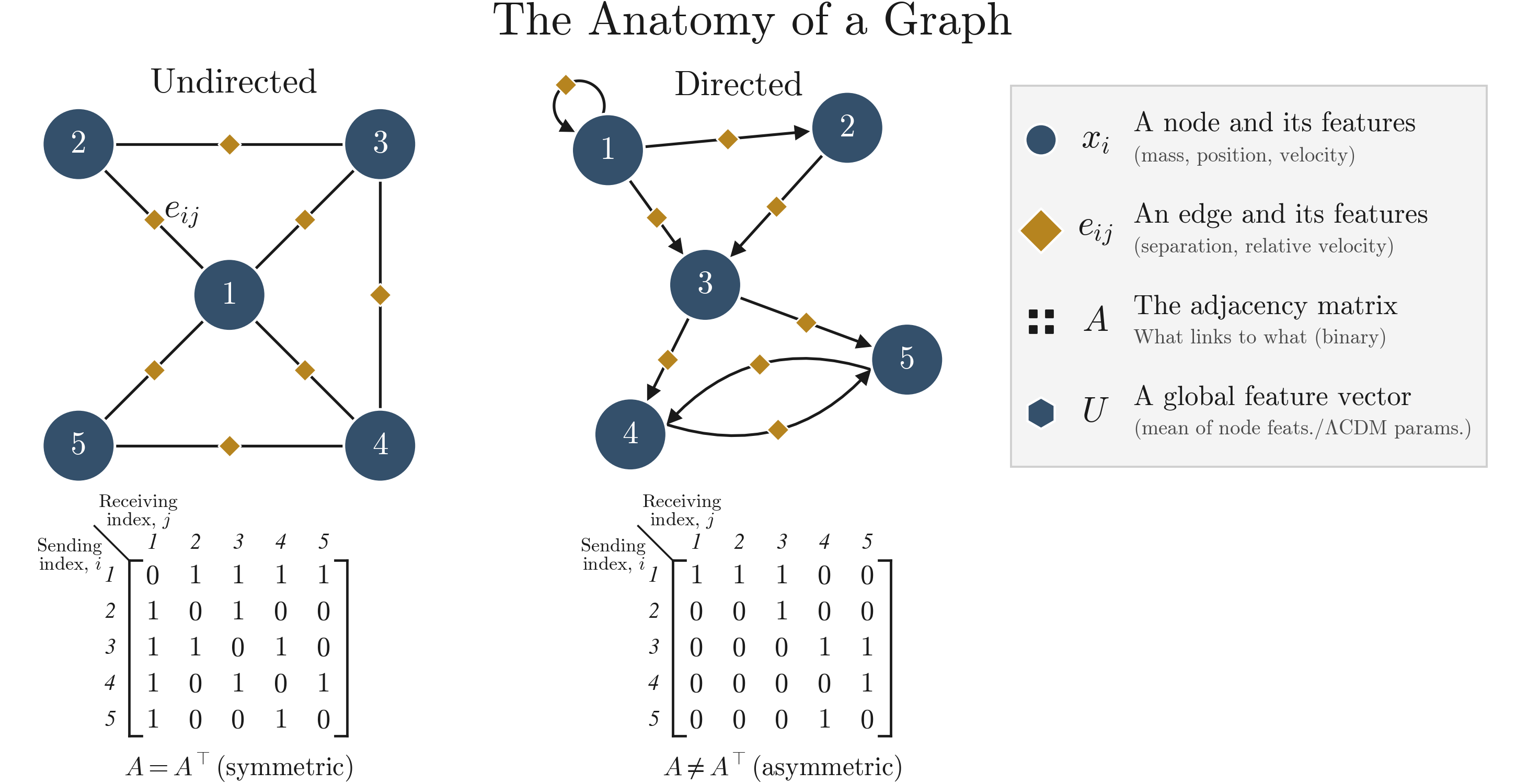}}
\caption{\textbf{The anatomy of a graph.} We write a graph as the triplet $G=(V,E,U)$: a set of nodes $V$, each carrying a feature vector $\mathbf{x}_i$ (here a halo's mass, position and velocity); a set of edges $E$, each potentially carrying a feature vector $\mathbf{e}_{ij}$ on the link between two nodes (here a separation and a relative velocity); and one global vector $U$ that speaks for the whole graph (e.g. cosmological parameters, or statistics pooled across the nodes). \emph{Left}, an undirected graph: the relations run both ways, as proximity, a bond, or a mutual pull, and the adjacency matrix stays symmetric, $A=A^{\top}$. \emph{Right}, a directed graph: the relation keeps a heading, as a merger, a flow, or the arrow of time, and the adjacency turns asymmetric, $A\neq A^{\top}$. The adjacency records what links to what, with $A_{ij}=1$ for an edge and $0$ otherwise, its rows the sending index $i$ and its columns the receiving index $j$. $A$ is useful as a complementary description of $E$, as it splits the pure graph topology from edge features. It also hands the graph to linear algebra through the matrix description. In the right graph, a \emph{self-loop} joins a node to itself, as a single entry on the diagonal, $A_{ii}=1$ (node~1). In message passing it lets a node keep a share of its own state while it folds in its neighbours'. Without it, round upon round of aggregation can sometimes wash a node's own identity away.}
\label{fig:graph_anatomy}
\end{figure}

A graph is either directed or undirected, and the physics of the relation should decide which. Directed edges suit the asymmetric, the causal, the time-ordered: a merger tree points forward in time, since to pass information backwards breaks causality. Undirected edges, a symmetric $A$, suit the symmetric relation: spatial proximity, a chemical bond, a pairwise gravitational pull. The \emph{neighbourhood} $\mathcal{N}(i)$ collects the nodes joined to $i$, and on it rests the whole notion of locality that follows. These structures are shown in Figure~\ref{fig:graph_anatomy}.

\subsection{Building the graph}
\label{subsec:no-adjacency}

Sometimes the graph arrives ready-made, its structure given by the data themselves; more often it does not, and we must decide --- using our physical intuition and reasoning --- who connects to whom. The question of graph creation in the two situations deserves separate treatment.

\subsubsection{When the graph structure is given a priori}
\label{subsubsec:graph-given}

\begin{figure}[t]
\centering
\includegraphics[trim={0cm 4.5cm 0 3.5cm},clip, width=0.7\textwidth]{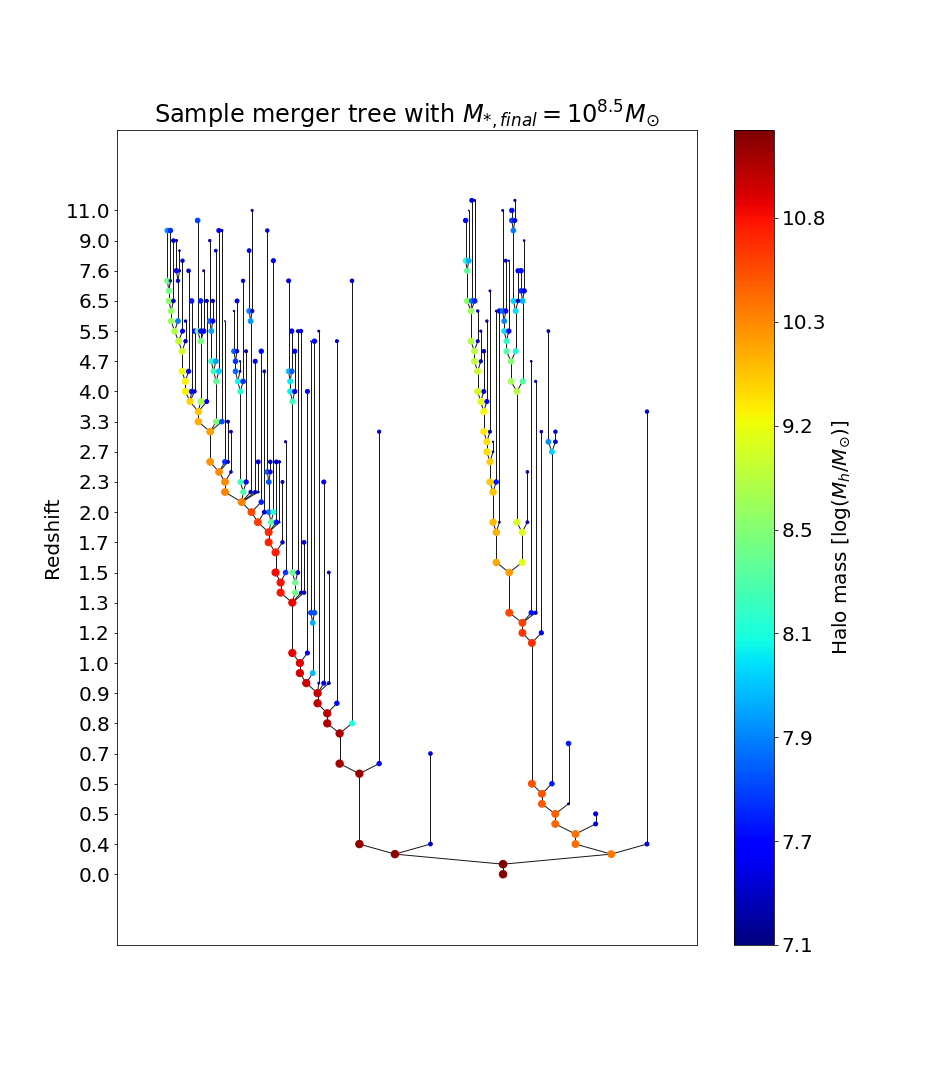}
\caption{A dark-matter halo merger tree: the assembly history forms a graph by construction, with nodes coloured by halo mass and edges that follow the flow of time. The directed edges make causality structural: information may flow from progenitor to descendant, but never back against the arrow of time. The merger tree is furthermore condensed so that only merger nodes are preserved, providing a strong inductive bias towards the importance of mergers. Adapted from \citet{Jespersen2022_mangrove}.}
\label{fig:merger_tree}
\end{figure}

When the data carry their own \textit{a priori} relations, the graph structure is given. This can be the case when the graph describes a causal process like a merger tree (Figure~\ref{fig:merger_tree}), or a detector with its wiring. As a first step, we need only a faithful transcription of the original structure, but we may still impose some inductive biases on our graphs.

As a case study, take the treatment of the merger trees in \citet{Jespersen2022_mangrove} and \citet{Chuang2024_mergertree_GNN}. The graph structure is inevitable and given \textit{a priori} in the dark-matter merger tree, where the halo assembly history forms a causal graph by construction (see Figure~\ref{fig:merger_tree}). However, we may choose to help our algorithms learn by imposing an \emph{inductive bias} on the algorithm through a restructuring of the graphs. The key insight is that since galaxies evolve smoothly when they are not merging, non-merging nodes are less important than merger nodes, an inductive bias that we can encode through the structure of the graph. \citet{Jespersen2022_mangrove} and \citet{Chuang2024_mergertree_GNN} have two different but related methods to solve this. \citet{Jespersen2022_mangrove} collapses the graph between mergers, keeping only nodes around mergers, thus leaving the GNN no choice but to focus on the mergers. \citet{Chuang2024_mergertree_GNN} keeps the snapshots between mergers, but in order to inductively bias the GNN towards learning the importance of mergers, adds edges between every merger node and all subsequent descendant nodes until the next merger. This direct connection between mergers and all descendant nodes acts as a strong and efficient inductive bias for learning their importance from our merger trees.

\subsubsection{When we must choose the graph structure}
\label{subsubsec:graph-chosen}

The natural world often presents us with point-cloud structured data. However, the connections between these points are rarely given \textit{a priori}. We may instead \textit{decide} which nodes connect to which. There are broadly four standard moves that may provide a way of \textit{choosing} the graph structure. We may \emph{choose} an edge set from physical knowledge:
\begin{itemize}
    \item Firstly, we may choose not to impose any graph structure. We then treat the data as a bare set and reach for a \emph{deep set}, the edgeless limit.\footnote{A set holds no native sense of locality: with no edges, no element lies near any other, so one cannot define a non-trivial coarsening on set structure alone. To coarsen, one must first infer a topology over the points, which precisely turns the set back into a graph.}
    \item We may also assume a \emph{fully connected} graph, maximally expressive but quadratic in cost. This choice can be reasonable for a small number of nodes, but for most applications, it is unreasonable.\footnote{A transformer is equivalent to a GNN on a fully connected graph, but much more computationally optimised.}
    \item We may attempt to \emph{learn} the adjacency jointly with the task, through neural relational inference \citep{Kipf2018_neural_relational_inference_GNN}, differentiable graph modules \citep{Kazi2023_DifferentiableGraphStructure}, or edge convolutions that rebuild a k-NN graph in feature space at every layer \citep{Wang2019_EdgeConv}. This has been done by \citet{Orsoe2025_graphnet2_library}, but is otherwise rare in the astrophysical literature.
    \item Finally, and most importantly, we may impose a \textit{physically or empirically motivated} graph structure using our hard-earned knowledge about the objects we study.
\end{itemize}

For the physicist the last option deserves repeated emphasis: we can often impose the graph from what we already know. For example, at a neutrino observatory, where every detector (node) registers when it received a hit, we may connect only the hits whose spatial separation stays consistent with a signal at or below the speed of light, a causal constraint no edge may violate. Mathematically, we express this as $A_{ij} = 1$ \texttt{if} $d_{ij} \leq c \cdot (t_j-t_i)$, where $d_{ij}$ is the distance between detector $i$ and detector $j$, and $t_i$ is the time at which detector $i$ registers a signal.

The neutrino observatory is of special interest for exploring the different options for how to construct a graph structure, since one setting admits all four possible approaches: to assume the light-cone structure to impose a physically motivated graph structure, to learn the adjacency matrix, to treat the hits as an unconnected point cloud, or to simply connect everything to everything. The literature has tried each of these, with details in \citet{Jespersen2021_icecube_bachelor, Bukhari2023_icecube_kaggle, Sogaard2023_graphnet_neutrino, Orsoe2025_graphnet2_library, Orsoe2026_nubench_benchmark}.

\begin{figure}[t]
\makebox[\linewidth][c]{
\includegraphics[trim={0cm 0.5cm 0 0},clip, width=1.25\textwidth]{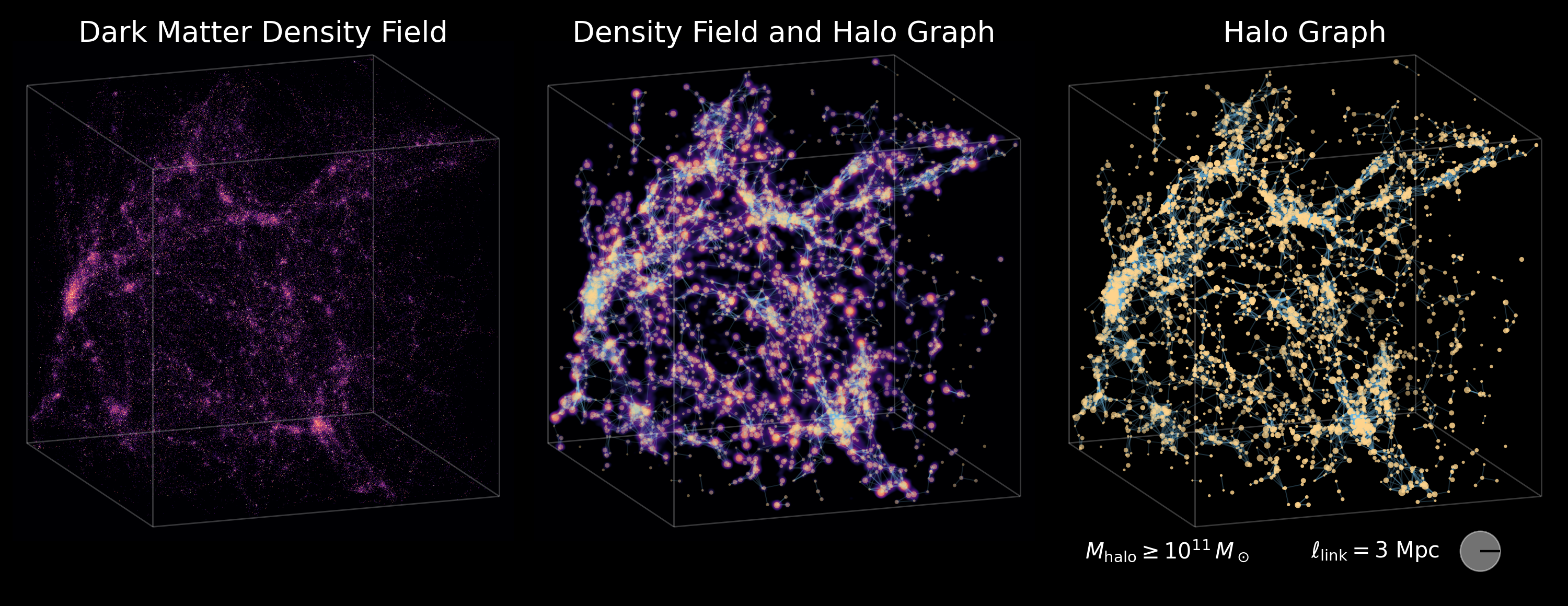}}
\caption{\textbf{The cosmic web as a proximity graph} over halos. Here the graph joins halos above $10^{11}\,M_\odot$ within a linking length of $3$ Mpc. k-NN and Delaunay constructions offer alternatives, each with its own inductive bias (\S\ref{subsec:no-adjacency}).}
\label{fig:cosmic_web}
\end{figure}

Another clear demonstration is the case of the cosmic web (Figure~\ref{fig:cosmic_web}). The halos or galaxies arrive as a point cloud, and we can then impose a family of different proximity graphs. Each choice answers the same question differently: what does it mean for two halos to be neighbours? The \emph{radius graph} fixes the neighbourhood scale: it connects every pair closer than some length $r$, often called the linking length \citep{Makinen2022_cosmicgraph_LSS, VillanuevaDomingo2022_cosmicgraph_clustering, Wu2023_galaxyenvironment_GNN, Wu2024_galaxyhalo_environment}. The implied bias holds that interactions are important below a certain physical scale: direct exchange reaches to $r$ and no further, correlations beyond $r$ must accumulate over successive layers. Furthermore, the degree of each node counts its neighbours within a fixed volume, so the connectivity itself records the local density. The \emph{k-NN graph} fixes the degree instead: it connects each object to its $k$ nearest neighbours, and neighbourhood scale now changes with the environment, short in the cluster, long in the void. The implied bias holds that rank matters more than distance: every halo receives the same number of messages and the same compute, no halo sits isolated however empty its surroundings. The local density thus vanishes from the connectivity as the k-NN graph fixes a homogeneous connectivity by construction. It will have to re-enter through the node features if it carries some desirable signal --- this is often the case. The \emph{Delaunay graph} removes these tuning knobs altogether: tessellate space on the points themselves and let the tessellation dictate the edges, the route \citet{Kololgi2025_cosmicweb_classification} take to classify cosmic-web environments. The construction carries no intrinsic neighbourhood scale, since the point set supplies its own everywhere, with short edges inside the cluster and long edges that bridge the void.

Beyond these proximity graphs stand the limits met in the list above: the deep set severs every halo from its surroundings, so the environment must arrive as a hand-built node feature or not at all, while the complete graph connects all pairs. In the case of the galaxies and halos, the latter will inevitably have to rediscover locality at an enormous computational cost. To include our prior knowledge saves us significant time and effort. Each construction states a physical belief in advance: the radius graph believes in a universal scale, the k-NN graph in the nearest company, the Delaunay graph in the geometry of local neighbourhoods of the point cloud itself. Since no belief clearly wins \textit{a priori}, we must treat the construction as a hyperparameter and explore it empirically. This then becomes a measurement in its own right, since the dependence of performance on, for example, linking length, reveals which environmental scales dominate the physics (see Figure~\ref{fig:linking_length}).

\section{Graph Neural Networks}
\label{sec:gnn}

A Graph Neural Network rests on one repeated idea: pass the message along the edge, aggregate, update the node. We open with the anatomy of a single layer and how it can be intuitively understood as a physical interaction network, then catalogue the common layers, the tasks they serve, and the modules from which we assemble any of the rest. Throughout, we follow the ideas of the blueprint in \S\ref{subsec:blueprint} applied to the graph.

\subsection{The anatomy of a message passing layer}
\label{subsec:message-passing}

A single message passing layer does three things, and the physicist has done all three before. Each node first receives messages from its neighbours, produced by a learnable function ($\psi$). Next, a permutation-invariant neighbourhood aggregator ($\bigoplus$) pools them across the neighbourhood of the node ($\mathcal{N}(i)$). Finally, the node updates its state to $\mathbf{h}_i$\footnote{$\mathbf{h}$ is used to designate a hidden state, whereas $\mathbf{x}$ typically denotes the original input features.} using a learnable function ($\phi$) on its own features and the neighbourhood aggregate:

\begin{equation}
    \label{eq:first_mp}
  \mathbf{h}_i \;=\; \phi\!\Big(\mathbf{x}_i,\ \bigoplus_{j\in\mathcal{N}(i)}
        \psi(\mathbf{x}_i,\mathbf{x}_j,\mathbf{e}_{ij})\Big).
\end{equation}

The two learnable functions are typically represented by weight matrices $\mathbf{W}$ which can then efficiently act on the input feature vectors $\mathbf{x}$. Equation~\ref{eq:first_mp} shows the most general message passing layer. Message passing can then be constrained to more specific forms given the task at hand. This is the subject of the following section. At its simplest the layer can be boiled down to $\mathbf{x}_i' = \mathbf{W}_1\cdot\mathbf{x}_i + \mathbf{W}_2\cdot\sum_{j\in\mathcal{N}(i)}\mathbf{x}_j$. This is the original \texttt{SAGEConv} layer of the \texttt{GraphSAGE} architecture \citep{Hamilton2017_GraphSAGE_inductive}. The original \texttt{SAGEConv} layer was defined with the mean aggregator, but it can be swapped for any permutation-invariant aggregator ---\footnote{All em-dashes in this work are organic and grass-fed.} like a sum. The form looks familiar for a reason: it is similar to a gravitational $N$-body update. Step the velocity of body $i$ forward, and the contributions of all other masses simply superpose:
\[
  \mathbf{v}_i^{t+1} \;=\; \mathbf{v}_i^{t}
    + \Delta t \sum_{j\neq i}
      \frac{G\,m_j(\mathbf{x}_j^{t}-\mathbf{x}_i^{t})}
           {|\mathbf{x}_j^{t}-\mathbf{x}_i^{t}|^{3}} .
\]
The carry-forward term plays $\phi(\mathbf{x}_i,\cdot)$; the pairwise pull $G m_j(\mathbf{x}_j-\mathbf{x}_i)/r^3$ plays the edge message $\psi$; the sum over bodies plays the permutation-invariant aggregator. A GNN layer, then, amounts to a learnable, generalised force law plus a one-step integrator, which explains why message passing feels native to the physicist, and why the learned edge function so often can be read back as a physical force through interpretability techniques \citep{Cranmer2019_symbolicphysics_GNN, Lemos2023_orbitalmechanics_symbolic}.

In the words of \S\ref{subsec:blueprint}, Equation~\ref{eq:first_mp} is the equivariant block $B$, acting on the whole node set at once:
\[
\mathbf{h} \;=\; B(\mathbf{x}), \qquad B(\mathbf{x})_i \;=\; \phi\!\Big(\mathbf{x}_i,\ \bigoplus_{j\in\mathcal{N}(i)}\psi(\mathbf{x}_i,\mathbf{x}_j,\mathbf{e}_{ij})\Big).
\]
Relabel the nodes and the outputs relabel with them, since the aggregator ignores neighbour order, so $B(g.\mathbf{x})=g.B(\mathbf{x})$ for every permutation $g$. Stack these blocks with nonlinearities between them, add an optional coarsening $P$ and the optional invariant readout pooling $A$ if the task is graph-level (\S\ref{subsec:pooling}), and the composition $f = A \circ \sigma_J \circ B_J \circ \cdots \circ \sigma_1 \circ B_1$ of the blueprint is now assembled on the graph.

\subsection{Common propagation layers}
\label{subsec:flavours}

\begin{figure}[p]
\hspace{0.1cm}
\makebox[\linewidth][c]{\includegraphics[trim={-1cm 0cm 0cm 0.5cm},clip, width=1.25\linewidth]{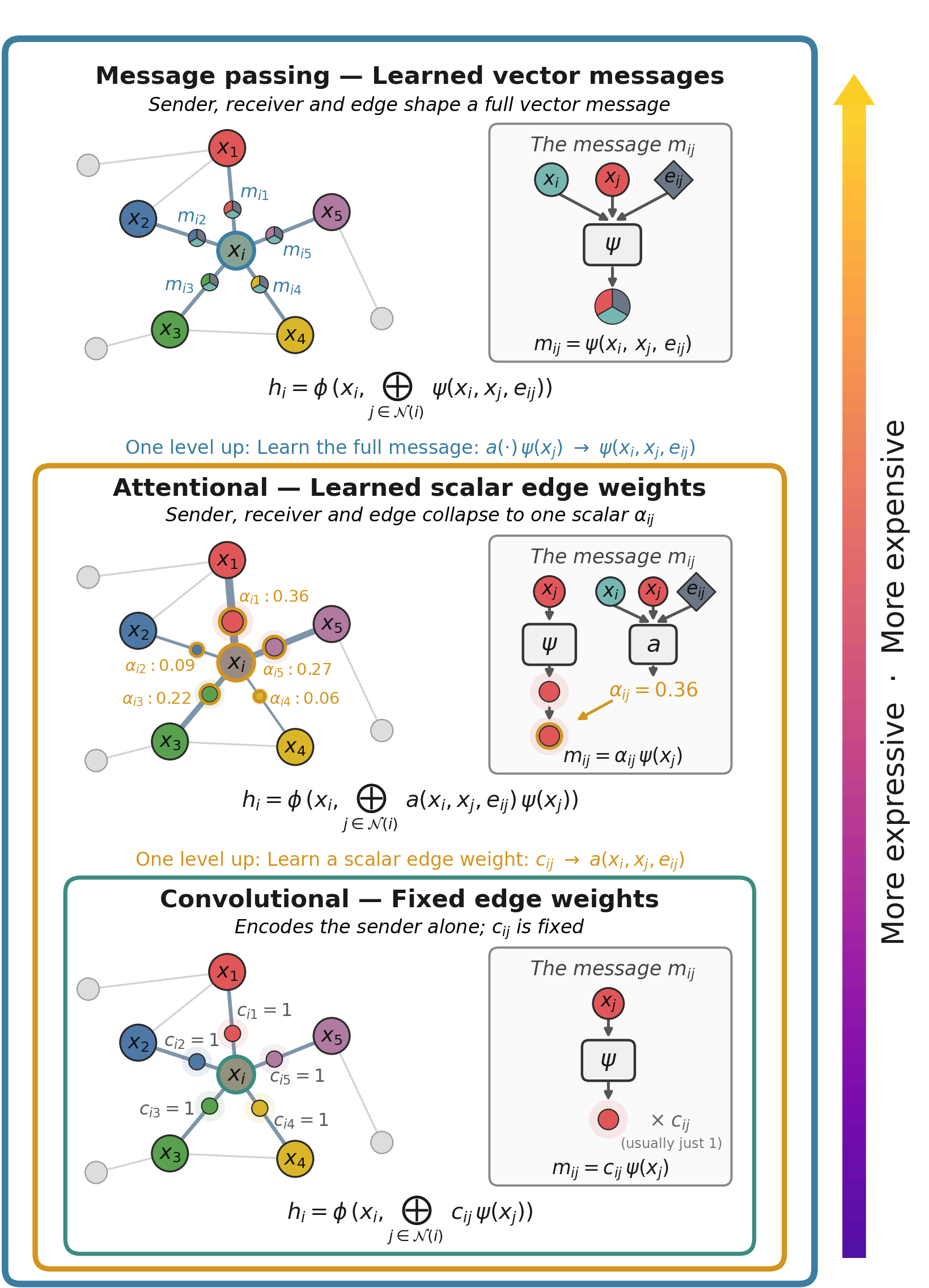}}
\caption{\textbf{The three canonical families of GNN layers}. The convolutional layer is a special case of the attentional layer, which in turn is a special case of the full message passing layer. As the degree of expressivity increases from convolutional to full message passing, the layer becomes harder and more expensive to train. We must always carefully consider this trade-off.}
\label{fig:flavours}
\end{figure}

The message passing layer defined in Equation~\ref{eq:first_mp} can be constrained and sorted into three canonical groups \citep{Bronstein2021_geometricDL_book, Velickovic2022_messagepassing_up}. In order of descending expressivity and computational cost:
\begin{align}
\text{Message passing:}\quad
  \mathbf{h}_i &= \phi\!\Big(\mathbf{x}_i, \textstyle\bigoplus_{j\in\mathcal{N}_i}
       \psi(\mathbf{x}_i,\mathbf{x}_j,\mathbf{e}_{ij})\Big),\\
\text{Attentional:}\quad
  \mathbf{h}_i &= \phi\!\Big(\mathbf{x}_i, \textstyle\bigoplus_{j\in\mathcal{N}_i}
       a(\mathbf{x}_i,\mathbf{x}_j, \mathbf{e}_{ij})\cdot\psi(\mathbf{x}_j)\Big), \\
\text{Convolutional:}\quad
  \mathbf{h}_i &= \phi\!\Big(\mathbf{x}_i, \textstyle\bigoplus_{j\in\mathcal{N}_i}
       c_{ij}\cdot\psi(\mathbf{x}_j)\Big).
\end{align}
The convolutional layer fixes the neighbour weight $c_{ij}$ (GCN; \pcite{Kipf2017_GCN_semisupervised}); the attentional layer learns a scalar weight $a(\mathbf{x}_i,\mathbf{x}_j,\mathbf{e}_{ij})$ (GAT; \pcite{Velickovic2018_GAT_attention}); the message passing layer computes a full vector message per edge (MPNN). To ascend the list buys expressivity at the cost of efficiency and compute (Figure~\ref{fig:flavours}).

\emph{Expressivity} here means the range of functions the layer can represent. Its formal theory, the Weisfeiler--Lehman ceiling, matters more to computer science than to astrophysics, but the interested can find it in \hyperref[app:sym:wl]{Appendix~1}. The message passing picture suffices for everything that follows.

Because the inner workings of GNNs inherently resemble both communication in a social network and interaction in a physical system, it is helpful to illustrate the workings of each this way.

\textbf{GNN Flavours as Social Communication:} Imagine that we want to update our belief about a matter of politics to become better informed about current affairs. How would we go about doing so? Well, we want to take our current belief $x_i$, and pass it through some function $\phi$ and arrive at our improved opinion $h_i$. For the most informed opinion, this update is based not just on our current opinion, but also on the opinions of those around us --- neighbours in our social network. The way that others may choose to communicate that information to us can be captured as a specific GNN layer type.

\begin{itemize}
    \item \textbf{Convolutional}: This layer constructs messages from neighbouring nodes based purely on the current features of those nodes. This is equivalent to people around us expressing their opinions without regard to the relation between us or to our own opinion. It is the GNN equivalent of non-existent social skills: the messages are passed regardless of the nature of our relationship and whether we may be offended.
    \item \textbf{Attentional}: This layer also constructs messages from neighbouring nodes based purely on the current features of those nodes, \textit{but} with the possibility of scaling those messages based on both the edge features and the features of the receiving node. The message itself is not altered, only its magnitude. This is equivalent to people around us expressing their unfiltered opinions without regard to the relation between us, but modulating the loudness of that opinion based on our relationship, and their knowledge of our current opinion.
    \item \textbf{Full message passing}: This layer constructs full messages taking in the features of sending and receiving nodes, as well as the edge features. This is equivalent to people around us changing the expression of their political opinion by taking into account the relation between us, and their knowledge of our current opinion. If they know that we disagree, they may choose to voice their opinion differently.
\end{itemize}

\textbf{GNN Flavours as Physical Laws:} Physical systems perform the same update: the state $x_i$ of an object passes through an interaction with its neighbours and arrives at the new state $h_i$. We select the layer type according to which information the interaction may access. Here, we give some examples of physical systems carrying the structures of each layer:
\begin{itemize}
    \item \textbf{Convolutional}: Temporal graphs encoding halo growth (merger trees) neatly represent this layer (see Figure~\ref{fig:merger_tree}). The descendant halo assembles as the unweighted sum over its progenitors --- fixed time sampling fixes every edge weight to unity, and the arrow of time bars the receiver from changing the message. The update of the descendant is set only by the current state and what it inherits directly from the past. What merging does to halo spin and concentration, for example, hides inside $\phi$.
    \item \textbf{Attentional}: Picture two pendulums on a shared beam. The first swings and its motion becomes the message. The beam, acting as our edge, carries the vibration across, and the beam's stiffness and damping (the edge features $\mathbf{e}_{ij}$) impact the degree to which the message survives the crossing. How strongly the second oscillator answers also depends on its own tuning. Driven at resonance it easily swings, but the same vibration simply passes a detuned oscillator by. The attention weight $a(\mathbf{x}_i, \mathbf{x}_j, \mathbf{e}_{ij})$ is exactly this uptake, set jointly by the sender's swing, the beam (edge) between them, and the receiver's own natural frequency. A node with many neighbours rests on many beams and listens selectively, amplifying the neighbours it is tuned to and damping the rest. The message is \textit{rescaled}, never \textit{rewritten}.
    \item \textbf{Full message passing}: The tidal interaction between two galaxies realises the full message passing layer. The perturbation depends on the mass and structure of each galaxy, and on the orbit that connects them: the prograde encounter tears out a tidal tail where the retrograde encounter at the same separation leaves the disc nearly untouched. No scalar weight can ever rescale one outcome into the other; the message itself must change form.
\end{itemize}

These can optionally take in a third kind of input besides the node and edge features: a global feature vector, $U$, with relevance for all nodes and edges. For a halo-related task, this could be the set of cosmological parameters governing the entire Universe, with shared importance for all halos. How a message passing layer composes all these kinds of inputs is shown in Figure~\ref{fig:full_interaction_network} and discussed more deeply in \S\ref{subsec:compose}.

Like other deep-learning architectures, GNNs train end-to-end by gradient descent on any differentiable loss function. These may be attached to the node-, edge-, or graph-level output with the loss function as e.g. cross-entropy when the label is a class, or mean squared error when it is a number. The loss also decides what the answer means. Swap the mean squared error for a Gaussian likelihood with a predicted variance, and the same network returns an error bar with every estimate. To gain a full posterior, the GNN can serve as the trainable summariser inside a normalising flow or a simulation-based inference pipeline, as in the void-finding flow of \citet{Thiele2026_void_graphflow}. We flag this role here and leave the machinery to the dedicated review of \citet{Thiele2026_SBI_review}.

\subsection{Prediction types}
\label{subsec:tasks}

\begin{figure}[t]
\includegraphics[width=\textwidth]{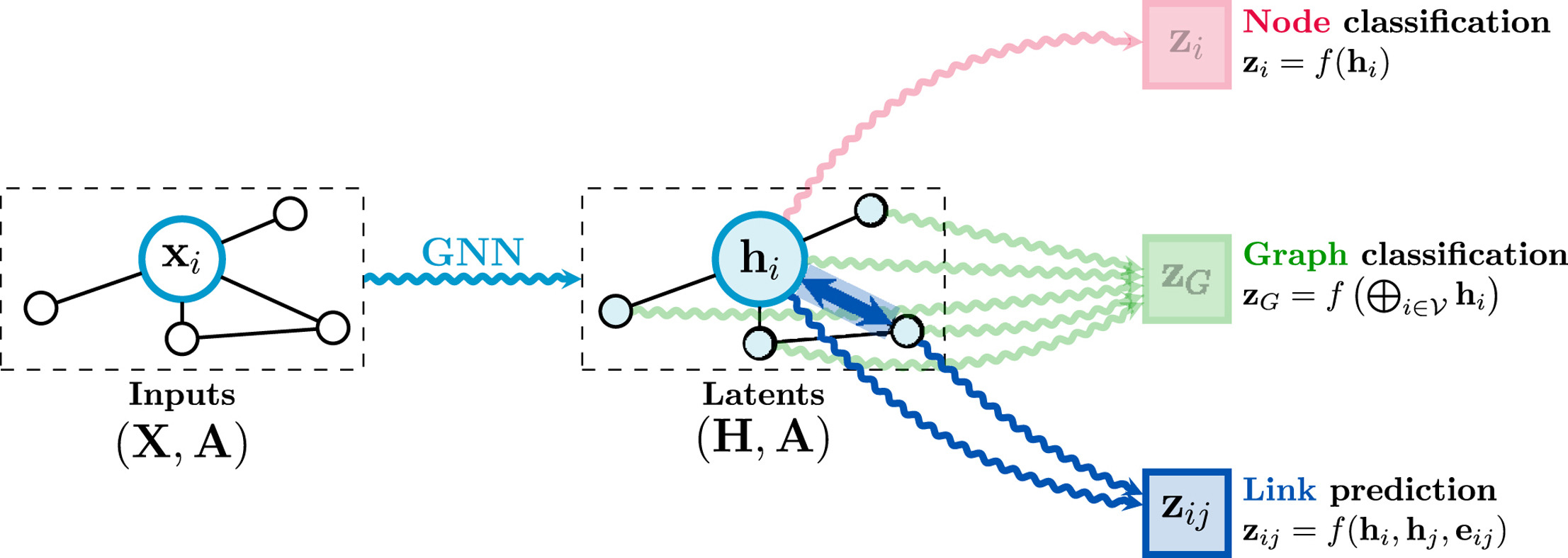}
\caption{\textbf{GNNs can be used across node-, edge-, and graph-level tasks}, for both classification and regression (the figure annotates the classification case). This versatility makes GNNs useful across most of astrophysics and astronomy. Figure reproduced after \citet{Velickovic2023_everythingconnected_review} with permission.}
\label{fig:tasks}
\end{figure}

A GNN affords every task one might want, across all three levels and both modes: the node, the edge, and the graph, in classification or regression mode (Figure~\ref{fig:tasks}). This adaptability makes GNNs a reasonable choice for the vast majority of astrophysical tasks, if not the ideal choice. On the cosmic web alone one may ask, at the graph level, whether the volume comes from a universe beyond $\Lambda$CDM, or what value $\Omega_m$ takes; at the node level, whether the galaxy in the halo is quiescent, or what its stellar mass is; at the edge level, whether two halos will merge, and in how long (see Figure~\ref{fig:cosmic_web}). The node- and edge-level outputs demand permutation \emph{equivariance} since the outputs must follow any permutations in the node ordering; the graph-level output demands permutation \emph{invariance}, since no relabelling of nodes should matter. The former is achieved with permutation-invariant aggregations when performing message passing, and the latter through a global permutation-invariant aggregation function over all nodes (see \S\ref{subsec:pooling}).

One more distinction matters before the tasks meet data: whether the network must generalise across many graphs, or only fill in the gaps of one. The literature calls the two modes inductive and transductive learning \citep{Hamilton2017_representation_review}.
\ifarxiv\enlargethispage{\baselineskip}\else\enlargethispage{5\baselineskip}\fi

\begin{svgraybox}
\textbf{Inductive vs.\ transductive learning.} Inductive learning\footnotemark{} trains on many graphs and deploys on new, unseen ones. Transductive learning trains and predicts within one fixed graph in order to spread labels from the nodes that have them to the nodes that lack them.
\end{svgraybox}
\footnotetext{The term carries an unfortunate near-collision: an inductive \emph{bias} (\S\ref{sec:intro}) is an assumption built into the model, while inductive \emph{learning} is a way of deploying one.}

The same distinction can be used to sort the astrophysical literature. The sorting is lopsided: nearly every application in \S\ref{sec:applications} is inductive, trained on many simulated merger trees, cosmic-web boxes, or detector events and deployed on new ones. The transductive setting still matters, as it looks like a familiar situation: one observed catalogue, partially labelled, as when a fraction of Milky Way stars carries spectroscopic follow-up and the rest awaits with only photometry. In the transductive setting the training and test nodes share edges by construction, so information leaks between the splits by default, a possible pitfall we return to in \S\ref{subsec:bp-diagnostics}.

\subsection{Composing edge, node, and global features into one layer}
\label{subsec:compose}

\begin{figure}[t]
\includegraphics[trim={0cm 14.65cm 0cm 0cm}, clip, width=\textwidth]{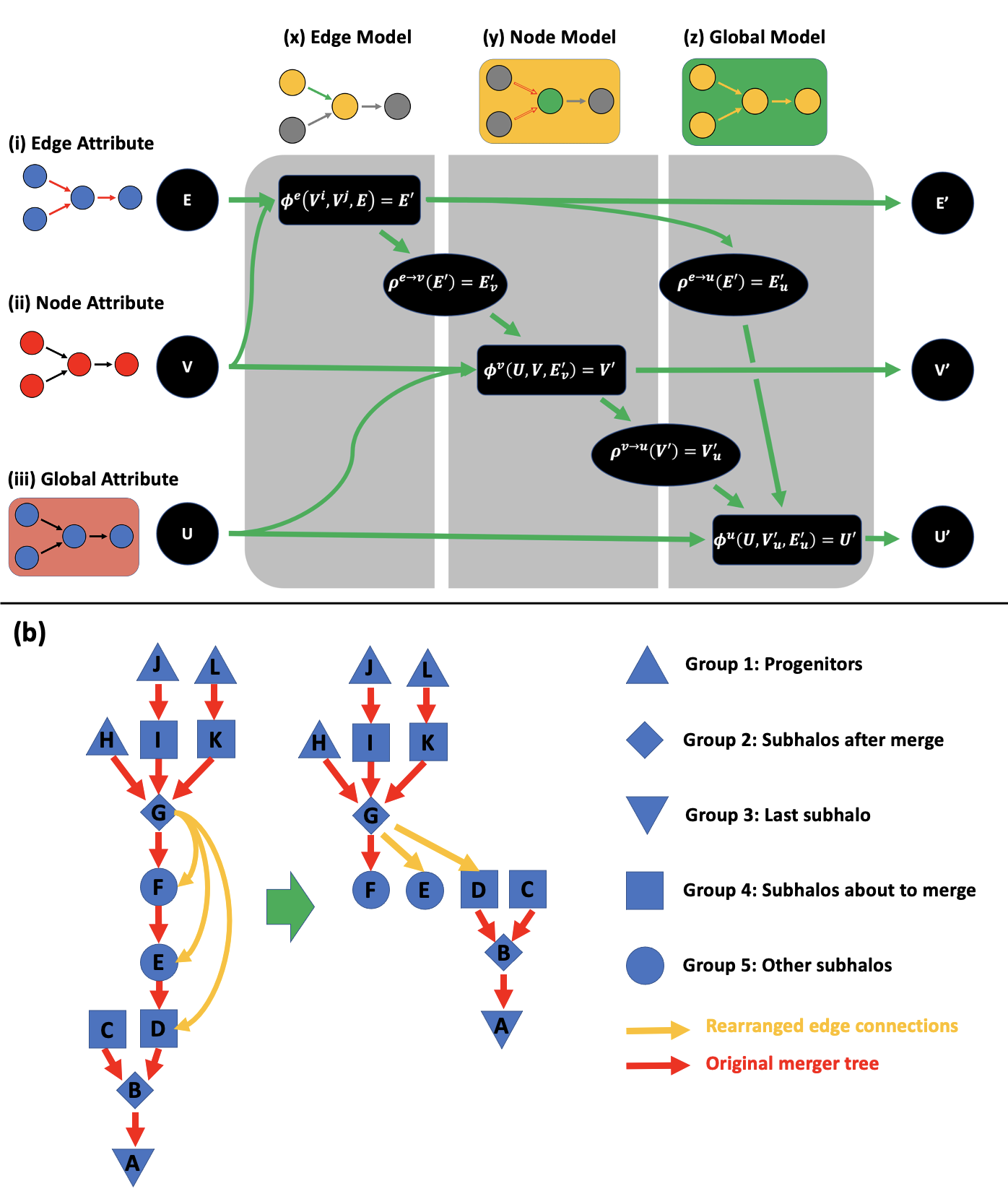}
\caption{\textbf{How to use edge, node, and global features in one graph-network layer}: the edge updates from its endpoints, the node from its incident edges, and the global feature from the whole graph. Figure reproduced from \citet{Chuang2024_mergertree_GNN}, itself closely inspired by the graph-network block of \citet{Battaglia2018_relational_GN}. The notation differs slightly from the one used here, but the compositions should still be clear.}
\label{fig:full_interaction_network}
\end{figure}

The three feature types, the edge $E$, the node $V$, and the global $U$, need not act in isolation: a full layer updates each in turn and lets information flow between them (Figure~\ref{fig:full_interaction_network}). One updates the edge from its endpoints, then the node from its aggregated edges, then the global feature from the whole. This order follows convention, not law, yet it composes the relational, the local, and the global into one permutation-respecting operation. The full structure is inspired by the \textit{Interaction Network} of \citet{Battaglia2016_interaction_network}, as generalised by \citet{Battaglia2018_relational_GN}.

\subsection{Invariant outputs through aggregations and poolings}
\label{subsec:pooling}

Two kinds of aggregations and poolings\footnote{These terms can be used interchangeably.} commonly take place in GNNs. The first aggregates the messages sent from neighbouring nodes; the second produces the graph-level output (as is so often the case in cosmology). In both, we must pool in a permutation-invariant way. The simplest pooling functions are the \texttt{sum}, \texttt{average}, \texttt{maximum}, and \texttt{minimum}. The sum grows with the node count and therefore retains it as a signal, while the mean stays invariant to it, a trade-off \S\ref{subsec:bp-choose} discusses further. Any of these functions, taken over the nodes, is order-independent, and so permutation-invariant.  Richer, distribution-aware pooling functions are rarer but do exist: the differentiable histogram, for one, captures higher moments and stays interpretable in the language of the classical set summary \citep{Cranmer2021_histpool_deepsets}. Hierarchical pooling instead coarsens the graph in stages \citep{Ying2018_diffpool_pooling}. In the end, we can understand the pooling/aggregation function as any statistically meaningful summary of the neighbourhood or graph: minimum, maximum, mean, variance, and onward.

It should be noted that any composition of permutation-invariant pooling functions itself constitutes a valid permutation-invariant pooling function. Indeed, it may often be beneficial to do so. We can, for example, concatenate the sum, variance, maximum, and minimum across the latent variables, resulting in a $4 \cdot N_\mathrm{lat}$-dimensional vector that can then be used for further calculation. To fully capture continuous features, it is in fact necessary to use multiple aggregators \citep{Corso2020_PrincipalNeighbourhoodAggregationGraph}. While this additional information about the latents is useful, we caution that the increased dimensionality can significantly complicate optimisation. The result of the trade-off is not obvious \textit{a priori}. However, if one is having trouble optimising a GNN, it is worthwhile trying to shrink or expand the number of aggregation functions.

\subsection{Symmetries in message passing}
\label{subsec:equivariant-mp}

Astrophysical signals are drenched in geometric structure: translations, rotations, and shears. Permutation symmetry, which every GNN respects by construction, says nothing about these, and a network fed raw data will happily learn functions that break the symmetries they are supposed to follow. However, there are many steps we can take to avoid this. To illustrate, we will look at rotation; the same three options apply for any symmetry.

To build the symmetry into the architecture is the first option. The $E(n)$-equivariant GNN of \citet{Satorras2021_EGNN_equivariant} keeps two ledgers per node and keeps them distinct: invariant scalar messages, functions of quantities such as pairwise distances, and equivariant vector updates, in which positions or velocities move only along difference vectors (see Figures~\ref{fig:rotation_demonstration} and \ref{fig:permutation_demonstration}). Every internal quantity then transforms correctly under $E(3)$, the group of rotations, translations, and reflections, by construction, and the guarantee holds at test time no matter what else the network learns. Note what $E(3)$ does and does not contain: a stretch or a shear breaks it, so a network built this way can still represent anisotropic physics like changes in clustering from e.g. redshift-space distortions. The intrinsic-alignment model of \citet{Jagvaral2022_intrinsicalignment_GNN} runs on exactly this principle.

While \citet{Satorras2021_EGNN_equivariant} remains a practical entry point for the astrophysical reader, the equivariant network has a rich family tree. Tensor-field networks and the SE(3)-Transformer encode features in spherical-harmonic channels \citep{Thomas2018_TensorFieldNetworks, Fuchs2020_SE3_GAT}, steerable\footnote{A \emph{steerable} feature is the network's version of a spherical tensor: under a rotation of the input, its components mix among themselves through a fixed matrix, the Wigner $D$-matrix of the corresponding order, and any two features couple through the physics-familiar Clebsch--Gordan coefficients \citep{CohenWelling2017_steerable_CNN}. The name comes from image processing, where a filter is steerable when its rotated copies are linear combinations of a fixed basis.} message passing extends the ledgers to arbitrary representations of the group \citep{Brandstetter2022_SteerableE3GNN}. The \texttt{e3nn} framework packages the machinery for the practitioner \citep{Geiger2022_E3NN_GeneralFramework}.

To feed the network nothing but invariants is the second option. Replace raw positions by relative separations and raw velocities by their moduli, and any architecture becomes rotation-invariant through its ``diet'' rather than its anatomy; this is the route of the field-level inference of \citet{deSanti2023_fieldlevel_LFI}, and much of its cross-simulation robustness travels through that choice. The option costs nothing but intellectual effort to implement.

To augment the training data with random rotations is the third option. It is often easiest to set up, but it is also the only ``solution'' that offers no guarantees. The network merely sees many orientations and is asked to infer the lesson. This option also carries significant extra compute cost, since more data are passed through the training procedure. When the symmetry is exact, architecture beats augmentation in both the strength of the guarantees and the cost of implementation. Equivariant GNNs for interatomic potentials reach target accuracy with three orders of magnitude fewer training examples than the non-equivariant counterpart \citep{Batzner2022_E3GNN_InteratomicPotentials}. On cosmological point clouds the benchmark of \citet{Balla2024_CosmicBenchmark_SymmetryPreservingData} shows that equivariant GNNs significantly outperform non-equivariant counterparts and domain-specific information extraction techniques in downstream performance as well as simulation efficiency, although the GNN struggles to fully use long-range information, a well-known issue we return to in \S\ref{sec:bestpractices}.

\subsection{Scaling up by sampling the neighbourhood}
\label{subsec:sampling}

When the graph grows to millions of nodes, we cannot pass every message everywhere all at once. Sampling controls the cost, by node, by layer, or by subgraph, and trades exactness for tractability and ease of optimisation \citep{Chiang2019_ClusterGCN, Zeng2020_GraphSAINT_graphsampling}. This may be a price worth paying when working at the scale of full galaxy surveys.

What typically limits us is the total amount of computation needed for a single message passing step. We can roughly think of the complexity of that computation as
\begin{equation}
    N_\mathrm{total} = N_\mathrm{E} \underbrace{\left[ \left( 2N_\mathrm{lat} + N_\mathrm{lat,edge} \right) N_\mathrm{hid} + N_\mathrm{hid} N_\mathrm{lat,out} \right]}_{\mathrm{Message}} + N_\mathrm{V} \underbrace{\left[ \left( N_\mathrm{lat} + N_\mathrm{lat,out} \right) N_\mathrm{hid} + N_\mathrm{hid} N_\mathrm{lat,out} \right]}_{\mathrm{Update}},
\end{equation}
where $N_\mathrm{E}$ and $N_\mathrm{V}$ count the edges and nodes of the graph, $N_\mathrm{lat}$, $N_\mathrm{lat,edge}$, and $N_\mathrm{lat,out}$ give the dimensions of the node, edge, and output latent vectors, and $N_\mathrm{hid}$ gives the width of the hidden layer. Since edges outnumber nodes in most graphs (note that a merger tree is an exception, as trees are among the sparsest graphs), the message term dominates, leading to the general scaling:
\begin{equation}
    N_\mathrm{total} \sim N_\mathrm{E} \, N_\mathrm{lat} \, N_\mathrm{hid}.
\end{equation}

Since the number of nodes often stays fixed while the mean degree $\langle k\rangle$ changes, it can be useful to rewrite this equation as
\begin{equation}
    N_\mathrm{total} \sim N_\mathrm{V} \langle k \rangle \, N_\mathrm{lat} \, N_\mathrm{hid}.
\end{equation}

As an example, a graph of $10^{5}$ nodes with $\langle k\rangle \sim 10^3$, some thirty features per node and ten per edge, trains to convergence in under an hour on a single modern GPU (an NVIDIA H100). Even with $N_\mathrm{total}$ on the order of billions, we can thus easily use our GNNs.

Two kinds of sampling have been used in the astrophysical literature:

\begin{itemize}
    \item \textbf{Neighbourhood Sampling}: Samples a fixed number of local neighbours per layer, thus constraining $\langle k\rangle$ to a constant. This kind of sampling is used in the original \texttt{GraphSAGE} implementation to limit memory overhead during network training,\footnote{This layer was developed to train on the full graph of all users of a social-media platform, so some simplifications were required.} and turned off during inference.
    \item \textbf{Subgraph/Cluster Sampling}: Splits up the graph into a fixed number of subgraphs, thus lowering $N_\mathrm{V}$. For example, \citet{Wu2023_galaxyenvironment_GNN} and \citet{Wu2024_galaxyhalo_environment}, investigating the dependence of galaxy property inference on environmental scale, used the \texttt{ClusterLoader} to make the training work as the environmental scale length (and thus the average number of edges) grows.
\end{itemize}

\subsection{The zoo of computational modules}
\label{subsec:zoo}

\begin{figure}[p]
\hspace{0.1cm}
\makebox[\linewidth][c]{\includegraphics[trim={0cm 1.25cm 0 0},clip, width=1.29\textwidth]{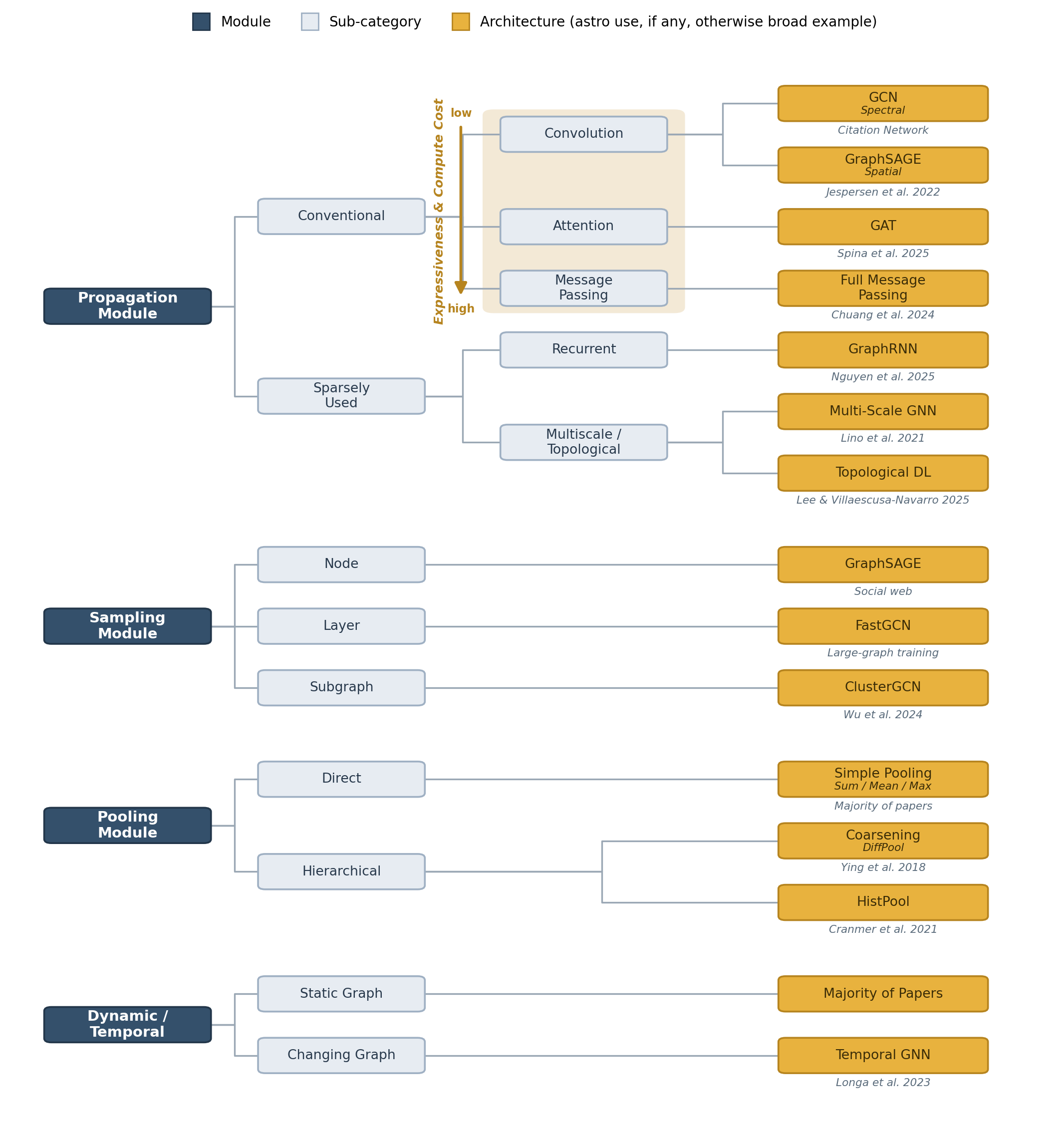}}
\caption{\textbf{A taxonomy of GNN computational modules}, restricted to an astrophysically useful subset, with a representative reference on each leaf. Adapted in the spirit of \citet{Zhou2020_GNN_review}.}
\label{fig:modules}
\end{figure}

The literature offers dozens of propagation layers, samplers, and pooling operators; \texttt{PyTorch Geometric} alone lists some sixty-five. Their number need not intimidate. Once one grasps the few principles behind the common layers, any of the rest sits a short step away; once more, the knowledge of certain principles supplies the knowledge of certain facts. Figure~\ref{fig:modules} organises an astrophysically useful subset as a taxonomy of computational modules.

\section{Applications in Astrophysics and Cosmology}
\label{sec:applications}

GNNs have spread across astrophysics wherever the data refuse to be organised on neat grids --- a very common occurrence. Below, we list and discuss the most prominent works using GNNs in astrophysics and cosmology. We organise the survey by subfield, and flag the divide between applications on real data (the minority) and the results that so far live only in simulation (the majority). The survey intends to illustrate and inspire; it does not exhaust. Table~\ref{tab:applications} collects it at a glance.

\newcommand{\redx}{\hspace{1pt}\mbox{\color{red}\ding{55}}\hspace{1pt}}
\newcommand{\greencheck}{\hspace{0.5pt}\mbox{\color{green!70!black}\ding{51}}\hspace{0.5pt}}

\begin{table}[p]
\raggedright
\hspace*{-1.2cm}\rotatebox{90}{%
\begin{minipage}{1.15\textheight}
\caption{The astro-GNN literature at a glance, organised by subfield and task. \emph{Level} gives the prediction type (\S\ref{subsec:tasks}): node, edge, or graph, with \emph{Mixed} where a subfield spans several. \emph{Graph construction} names how the graph was constructed (\S\ref{subsec:no-adjacency}). \emph{Layer} gives the propagation flavour of \S\ref{subsec:flavours}, where \emph{Mixed} usually covers convolutional and message passing. \emph{Real data} marks if any analysis has included real observations, still by far the minority of published works (see \S\ref{sec:frontier}). References are representative, not exhaustive.}
\label{tab:applications}
\centering
\footnotesize
\setlength{\tabcolsep}{4.5pt}
\resizebox{\textwidth}{!}{%
\begin{tabular}{p{2.2cm} p{5.1cm} p{1.05cm} p{2.4cm} p{2.2cm} p{1.05cm} p{5.3cm}}
\hline\noalign{\smallskip}
Subfield & Task & Level & Graph construction & Layer & Real data? & Representative references \\
\noalign{\smallskip}\svhline\noalign{\smallskip}
Cosmology/LSS & $\Omega_m$, $\sigma_8$ from halo/galaxy catalogues & Graph & Radius & Message passing & No (\redx) & \citeauthor{Makinen2022_cosmicgraph_LSS} (2022); \citeauthor{VillanuevaDomingo2022_cosmicgraph_clustering} (2022); \citeauthor{deSanti2023_fieldlevel_LFI} (2023) \\
 & Continuous-field reconstruction from tracers & Node & Radius & Message passing & No (\redx) & \citeauthor{Kvasiuk2024_fieldreconstruction_GNN} (2024); \citeauthor{Krywonos2026_GNNCNN_reconstructing_DM_from_galaxies} (2026) \\
 & Cosmic-web environment classification & Node & Delaunay/proximity & Attentional & No (\redx) & \citeauthor{Kololgi2025_cosmicweb_classification} (2025) \\
 & Multi-scale cosmology; benchmarks & Graph & Topological/multi-scale & Mixed & No (\redx) & \citeauthor{Lee2025_topological_cosmology} (2025); \citeauthor{Huang2025_CosmoBench_benchmark} (2025) \\
 & CMB, lensing, PNG, voids & Mixed & Sphere/proximity & Mixed & No (\redx) & \citeauthor{Adams2023_CMB_graphconv} (2023); \citeauthor{Park2023_lensing_bayesianGNN} (2023); \citeauthor{Jung2023_QuijotePNG_massfunction} (2023); \citeauthor{Thiele2026_void_graphflow} (2026) \\
\noalign{\smallskip}
Galaxies \& Halos & Galaxy properties from merger trees & Graph & Tree & Mixed & No (\redx) & \citeauthor{Jespersen2022_mangrove} (2022); \citeauthor{Chuang2024_mergertree_GNN} (2024) \\
 & Galaxy--halo connection from environment & Node & Radius & Message passing & No (\redx) & \citeauthor{Wu2023_galaxyenvironment_GNN} (2023); \citeauthor{Wu2024_galaxyhalo_environment} (2024) \\
 & Halo/cluster masses from neighbourhoods & Node & Radius & Message passing & Yes (\greencheck) & \citeauthor{VillanuevaDomingo2022_halomass_GNN} (2022); \citeauthor{VillanuevaDomingo2023_MW_mass} (2023) \\
 & Cluster masses from X-ray profiles & Graph & Sequence-as-graph & Attentional & Yes (\greencheck) & \citeauthor{Iqbal2025_cluster_xray_GNN} (2025) \\
 & Intrinsic alignments (generative) & Node & Radius & $E(3)$-equivariant & No (\redx) & \citeauthor{Jagvaral2022_intrinsicalignment_GNN} (2022) \\
 & Generative trees and environments & Graph & Tree/radius & Mixed & No (\redx) & \citeauthor{Tang2022_merger_normalizingflow} (2022); \citeauthor{Nguyen2024_FLORAH_assembly} (2024); \citeauthor{Nguyen2025_mergertree_generative} (2025); \citeauthor{Nguyen2026_DREAMS_diffusion} (2026) \\
\noalign{\smallskip}
Milky Way, Stars, Local Volume & Chemical/kinematic tagging & Node & Complete/learned & Attentional & Yes (\greencheck) & \citeauthor{Spina2025_chemicaltagging_GAT} (2025); \citeauthor{Berni2025_MWhalo_substructure} (2025) \\
 & Dwarf-galaxy dark-matter profiles & Graph & k-NN over stars & Message passing & Yes (\greencheck) & \citeauthor{Nguyen2023_dwarf_densityprofile} (2023); \citeauthor{Nguyen2025_dwarf_GraphNPE} (2025); \citeauthor{Nguyen2026graphNPE_applied} (2026) \\
\noalign{\smallskip}
Dynamics & Force laws and masses by symbolic regression & Edge & Complete & Message passing & Yes (\greencheck) & \citeauthor{Cranmer2020_symbolicregression_inductivebias} (2020); \citeauthor{Lemos2023_orbitalmechanics_symbolic} (2023) \\
\noalign{\smallskip}
Instruments & Fibre assignment, resource allocation & Node & Conflict graph & Message passing & Yes (\greencheck) & \citeauthor{Cranmer2021_resourceallocation_unsupervised} (2021); \citeauthor{Wang2022_spectroscopy_resourceallocation} (2022); \citeauthor{Sun2026_fiberassign_GNN} (2026) \\
 & Neutrino- and muon-event reconstruction & Graph & k-NN over hits & Mixed & Yes (\greencheck) & \citeauthor{Choma2018_IceCube_classification} (2018); \citeauthor{Abbasi2022_IceCube_lowenergy} (2022); \citeauthor{Sogaard2023_graphnet_neutrino} (2023); \citeauthor{Soldin2025_IceCube_GCN} (2025) \\
 & Cosmic rays, interferometry & Graph & Detector geometry & Mixed & Yes (\greencheck) & \citeauthor{Ferriere2025_GRAND_cosmicray} (2025); \citeauthor{Rodriguez2025_Auger_photon} (2025); \citeauthor{Kannan2025_interferometer_GNN} (2025) \\
\noalign{\smallskip}\hline
\end{tabular}}
\end{minipage}}
\end{table}

\subsection{Cosmology and Large-Scale Structure}
\label{subsec:app-cosmo}

Cosmology supplies the largest graphs and point clouds that astronomy as a field must work with. This is especially true of Large-Scale Structure (LSS), where large-scale surveys such as DESI now map out the precise position of $\sim10^{8}$ objects tracing the LSS. Inferring cosmological parameters from this enormous point cloud is usually done through N-point\footnote{With $N\in\{2,3\}$ and, to a lesser extent, $4$.} correlation functions that heavily compress the information content. Field-level inference methods are being developed to circumvent this limitation \citep{Bayer2023_MCLMC_fieldlevel}, but at present they still rely heavily on binning and smoothing. Traditional methods must furthermore apply significant scale cuts to keep the likelihood analytically tractable, discarding large amounts of information. Simulation-based inference is now being increasingly applied to circumvent the need for explicit likelihoods \citep{Hahn2023_SimBIG_forwardmodel, Hahn2024_SimBIG_constraints, Ho2024_LtU_ILI}, although inherent degeneracies in the parameter space are still present \citep{Krishnaraj2026_transferlearning}.

\begin{figure}[t]
\centering
\includegraphics[width=0.8\textwidth]{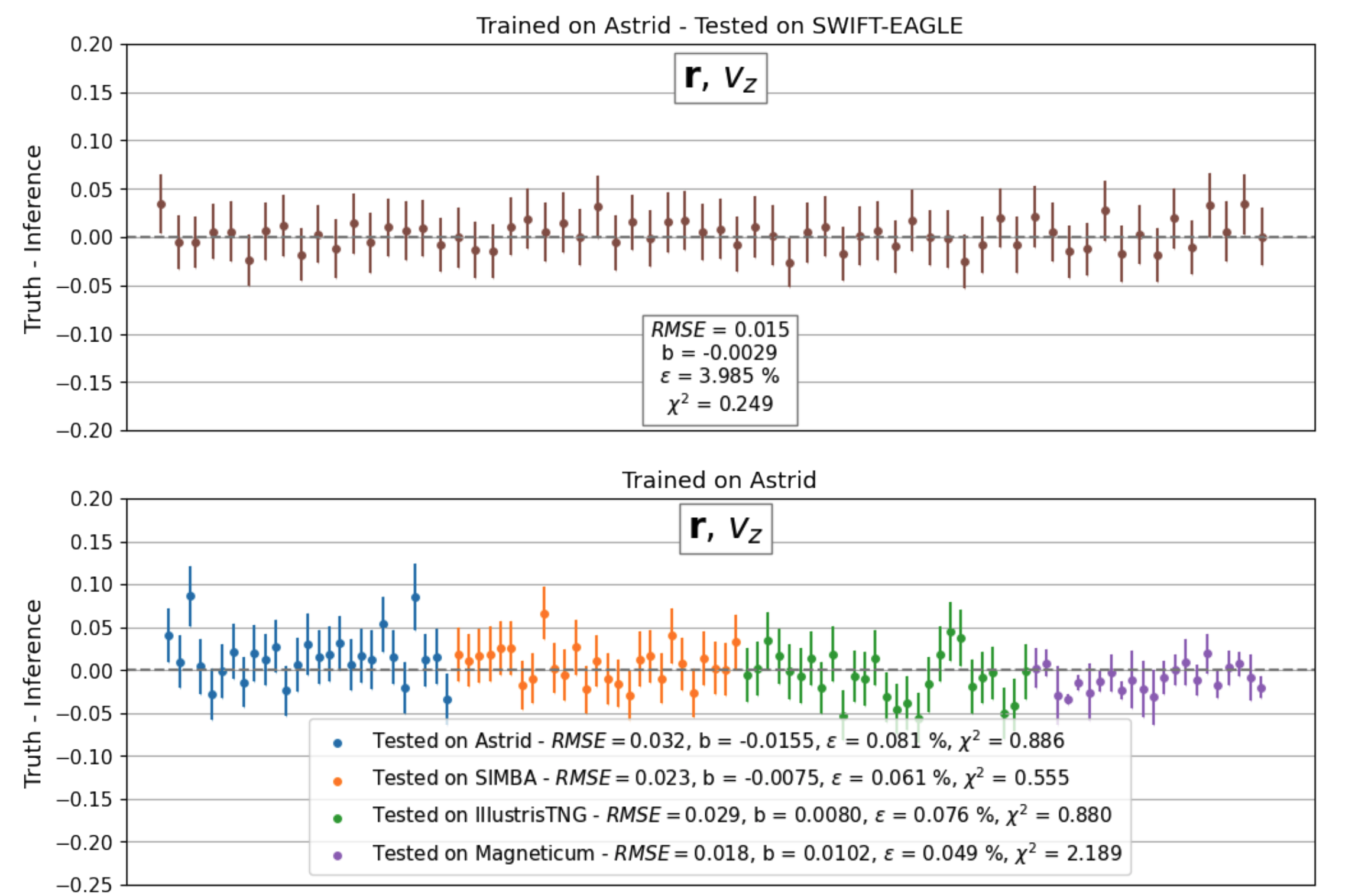}
\caption{\textbf{Robustness in cosmological parameter inference with GNNs}. Field-level inference of cosmological parameters survives a change of galaxy-formation model when a GNN is the backbone. Each panel shows the difference between the true and the inferred $\Omega_m$ for a GNN that reads only galaxy positions and line-of-sight velocities $(\mathbf{r}, v_z)$. Trained on the Astrid simulations, the network holds its performance on simulations it never saw: SWIFT-EAGLE (top) and the cosmic-variance sets of Astrid, SIMBA, IllustrisTNG, and Magneticum (bottom). The authors attribute the robustness to carefully invariantised inputs (\S\ref{subsec:equivariant-mp}, \S\ref{subsec:bp-bias}). Figure 2 of \citet{deSanti2023_fieldlevel_LFI}, reproduced from the published article at \href{https://iopscience.iop.org/article/10.3847/1538-4357/acd1e2}{doi:10.3847/1538-4357/acd1e2} (ApJ 952, 69), \copyright~2023 The Author(s), under the CC BY 4.0 licence.}
\label{fig:desanti_robustness}
\end{figure}

A GNN naturally mitigates both problems. Training on simulations renders the likelihood implicit, and the graph framing enables field-level inference with almost no information discarded. A GNN infers cosmological parameters straight from the catalogue: the cosmic graph over halos and galaxies has been shown to constrain $\Omega_m$ and $\sigma_8$ better than any standard summary statistic \citep{Makinen2022_cosmicgraph_LSS, VillanuevaDomingo2022_cosmicgraph_clustering}. While GNNs are especially strong, even just viewing the cosmic web as a graph and employing graph-theoretic analysis models can improve constraints on $\Omega_m$ and $\sigma_8$ \citep{Yang2023_haloclustering_graph}. These networks are furthermore more interpretable than other deep-learning methods \citep{Shao2023_omegam_universalequation}. More importantly, the robust heuristic that galaxy positions act as biased tracers of the LSS is embedded in the graphs, making field-level likelihood-free inference of cosmological parameters with GNNs robust to changes in galaxy formation physics \citep[][Figure~\ref{fig:desanti_robustness}]{deSanti2023_fieldlevel_LFI}. This also makes GNNs prime candidates for domain adaptation and robustness methods, since the modifications needed are small, a necessary condition for domain adaptation to be successful \citep{Roncoli2023_domainadaptation_cosmology, Lee2024_SDSS_domaingeneralization}. Note that \citet{Lee2024_SDSS_domaingeneralization} reaches the survey data with a deep-set-style point-cloud network (Minkowski-PointNet), the edgeless limit of \S\ref{subsec:no-adjacency}, rather than a GNN proper. Inference from these point clouds of discrete tracers can also provide improved maps to continuous fields like the baryon, dark matter, and electron density fields \citep{Kvasiuk2024_fieldreconstruction_GNN, Krywonos2026_GNNCNN_reconstructing_DM_from_galaxies}. GNNs are now also used to more efficiently and accurately sort the cosmic web into different environment types \citep{Kololgi2025_cosmicweb_classification}.

One of the challenges facing GNNs is to capture both long- and short-range dependencies at the same time while remaining computationally efficient. \citet{Lee2025_topological_cosmology} recently introduced a \textit{topological} version of the GNN that combines graphs on different scales to jointly capture small and large-scale interactions without growing the graph to contain billions of edges. The approach keeps the computation efficient while enabling multi-scale interaction.

Many other cosmological tasks are now also being solved with graph-based methods. This includes CMB recovery \citep{Adams2023_CMB_graphconv}, Bayesian lensing inference \citep{Park2023_lensing_bayesianGNN}, inference of primordial non-Gaussianity \citep{Jung2023_QuijotePNG_massfunction}, and the hunt for cosmic voids with graph flows \citep{Thiele2026_void_graphflow}. These diverse tasks fill out the field. Given this wide applicability and promise, common benchmarks for the use of GNNs in cosmology now appear \citep{Huang2025_CosmoBench_benchmark}.

\subsection{Galaxy formation and the galaxy--halo connection}
\label{subsec:app-galaxies}
The bond between a galaxy and its dark-matter halo is deep and extends in both time and space. The temporal dimension is captured in merger trees (Figure~\ref{fig:merger_tree}), and the spatial dimension is captured in local neighbourhood graphs.

\begin{figure}[t]
\centering
\includegraphics[width=0.65\textwidth]{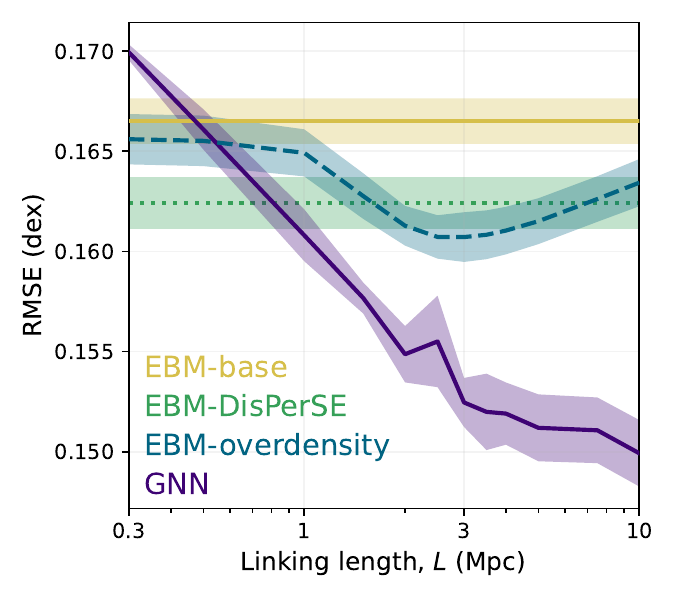}
\caption{\textbf{The environmental galaxy--halo connection is captured by the GNN}. The error (RMSE, dex) on predicted galaxy stellar mass against the linking length $L$ of the environment graph. A GNN applied to a halo proximity graph with changing linking length reveals how much information each scale contains. The explainable-boosting-machine baselines, fed scalar summaries of the environment (the overdensity, DisPerSE cosmic-web features), cannot reproduce the environment connection with a similar fidelity. Note that even at $L=10$\,Mpc, the edge of the tested range, the GNN error still falls. Information remains to be gained from still larger scales. Reproduced from \citet{Wu2024_galaxyhalo_environment}.}
\label{fig:linking_length}
\end{figure}

The galaxy--halo--environment connection is also captured through the use of GNNs, since small-scale clustering (see \S\ref{subsec:app-cosmo}) is naturally captured through locally connected graphs like the one shown in Figure~\ref{fig:cosmic_web}. These environment-GNNs can then learn accurate maps between baryonic galaxy properties and halo properties \citep{Wu2023_galaxyenvironment_GNN, Wu2024_galaxyhalo_environment}. Conventional environmental scalar summary statistics cannot make this mapping at the same level \citep{Wu2024_galaxyhalo_environment}, and even summary statistics directly constructed to capture the geometry of the cosmic web \citep{Sousbie2011_cosmicweb_theory} fall short of the GNN which learns it directly from data. Since these analyses must assume a linking length to define the scale of the environment, the performance of the GNN as a function of linking length also informs what environmental scales are most important for galaxy/halo evolution (Figure~\ref{fig:linking_length}, built on the graph of Figure~\ref{fig:cosmic_web}).

These environment-GNNs can then be used in observations to infer halo masses from observed neighbourhoods, a major advance towards inferring individual halo properties from observations \citep{VillanuevaDomingo2022_halomass_GNN, VillanuevaDomingo2023_MW_mass, Garuda2024_GNN_reconstructing_halomass_from_galaxies, Larson2024_halomass_images}. Halo triaxiality can also be inferred with GNNs \citep{Delgado2025_cluster_triaxiality}. Other galaxy quantities determined by local galaxy clustering are also constrained with GNNs. Intrinsic alignment, a vector quantity, is naturally captured by $E(3)$-equivariant GNNs \citep{Jagvaral2022_intrinsicalignment_GNN}.

One particularly creative use of a GNN came from \citet{Iqbal2025_cluster_xray_GNN}, who used the flexibility of the graph framing to train a single network to map from one-dimensional X-ray profiles to cluster masses. Different clusters have different numbers of observed points along their 1D X-ray profiles. The data therefore suit the flexibility of a GNN and its adaptive geometry. The graph neatly conserves the coupling and equilibrium between neighbouring bins, which is what is learned by the GNN and allows for robust mass inference. Crucially, having calibrated on hydrodynamical simulations, they apply the trained network to real galaxy clusters observed with XMM-Newton --- one of the few scientific (rather than instrumental) GNN analyses to reach real data at all.

One direction which is increasingly explored is \textit{generative modelling} of environments \citep{Nguyen2026_DREAMS_diffusion} and assembly histories \citep{Tang2022_merger_normalizingflow, Nguyen2024_FLORAH_assembly, Nguyen2025_mergertree_generative}. Although these generative models are impressive, note that the generation can only be done under significant assumptions to constrain the generative process, an issue common to all generative graph tasks. This is discussed further in \S\ref{sec:frontier}.

\subsection{Stellar populations, the Milky Way, and dark-matter substructure}
\label{subsec:app-stellar}
Closer to home, there is also ample geometric structure to embed in our learning algorithms. GNNs can sort stars and probe the dark matter substructure from stellar kinematics and satellite populations. Chemical and kinematic tagging, a method to reunite stars with a common origin, can be recast as a graph problem with graph attention applied to a graph with chemical abundances as node features, and orbital similarity on the edges \citep{Spina2025_chemicaltagging_GAT}. Substructure in the Milky Way halo (streams, globular clusters, and merger remnants) also emerges from applying GNNs to Gaia and APOGEE \citep{Berni2025_MWhalo_substructure}. Graph kernels can also be used for these tasks, although the reader should note that using a graph kernel does not make the resulting network a GNN \citep{Ting2023_WLkernel_chemicaltagging}. The density profiles of galaxies, a direct handle on the nature and physics of dark matter, are directly inferable with graph networks applied to stellar kinematics \citep{Nguyen2023_dwarf_densityprofile, Nguyen2025_dwarf_GraphNPE}, a method that has recently been applied observationally to dwarf galaxies \citep{Nguyen2026graphNPE_applied}.

\subsection{Dynamics}
\label{subsec:app-dynamics}

\begin{figure}[t]
\centering
\includegraphics[width=0.8\textwidth]{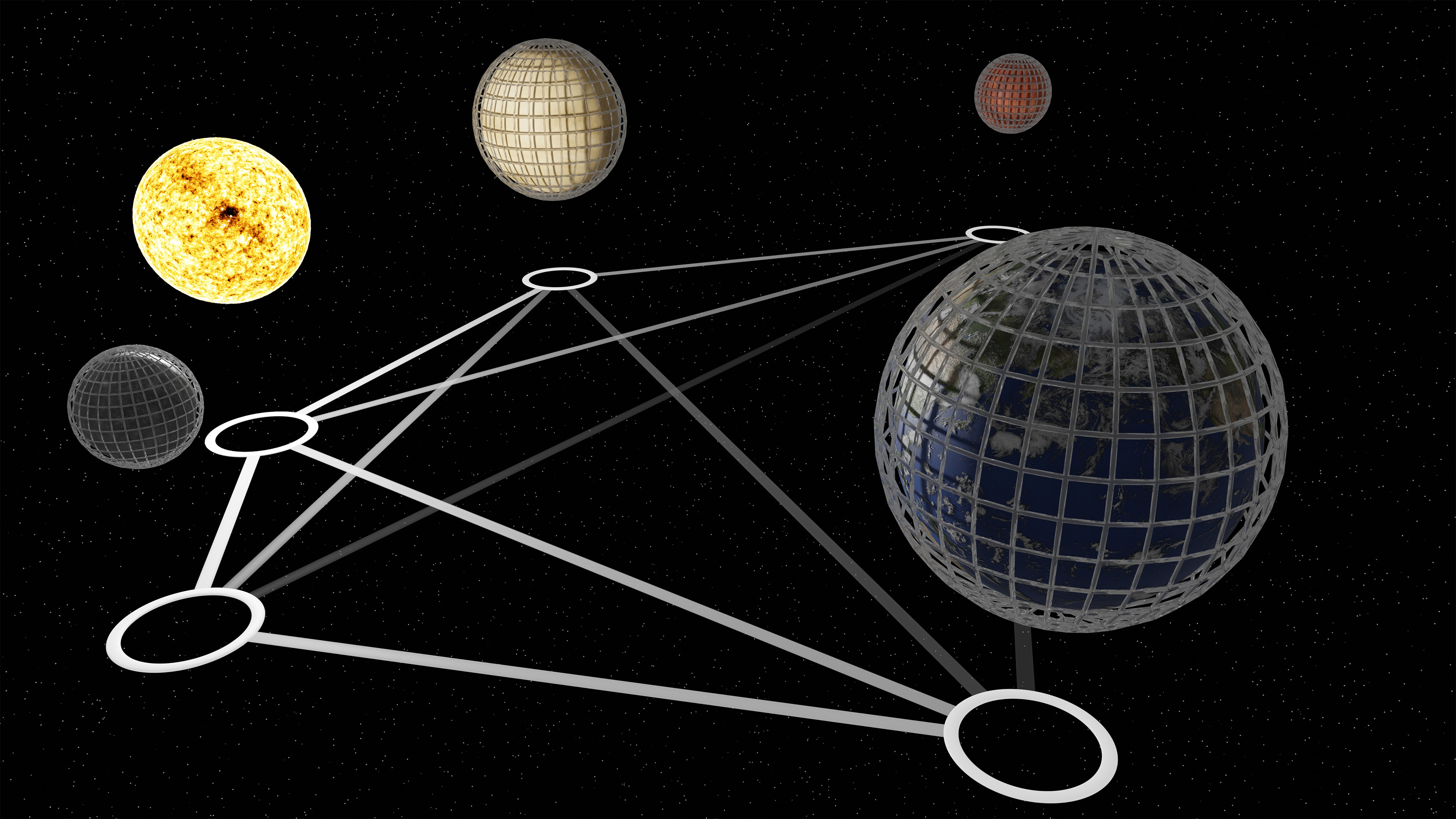}
\caption{\textbf{The solar system embedded as a graph}. The Sun, the planets, Pluto, and the largest moons form the nodes, drawn from thirty years of NASA HORIZONS ephemerides sampled every thirty minutes; the messages learn the pairwise gravitational interaction. Trained on these orbits, the interaction network learns the dynamics, and symbolic regression over its messages gives us back Newton's law of gravitation together with the masses of the bodies. Reproduced from \citet{Lemos2023_orbitalmechanics_symbolic} under the CC BY 4.0 licence.}
\label{fig:solar_system_graph}
\end{figure}

Dynamics are governed by forces enacted by particles on one another within some background potential. The GNN turns the force-law analogy of message passing presented in \S\ref{subsec:message-passing} literal. The interaction network naturally learns the equations of motion of the $N$-body system and, combined with a sparsity constraint on the dimensionality of the learned latent messages, symbolic regression can hand messages back to us as a readable law \citep[see e.g. the solar system in Figure~\ref{fig:solar_system_graph}, as well as the work of][]{Cranmer2019_symbolicphysics_GNN, Cranmer2020_symbolicregression_inductivebias, Lemos2023_orbitalmechanics_symbolic}. This is due to the inherent interpretability of vector messages as forces, another dimension of the physics-like nature of GNNs. Related deep-learning machinery predicts the stability of the compact planetary system \citep{Cranmer2021_planetarystability_BNN}. The broader programme of the learned simulator on irregular meshes with voxels locally coupled to one another \citep{SanchezGonzalez2020_GNS_physics, Pfaff2021_meshsim_GNN} awaits its turn to be applied in astrophysics in earnest, although a few early project proposals in this direction are underway.

\subsection{Observation scheduling, detectors, and instrumentation}
\label{subsec:app-instruments}

Point a telescope towards the sky, and the combinatorial and geometric trouble starts before the first photon --- how do we select which observations to take from billions of possibilities, and when exactly? And once we have our observations spread out across our detectors, how do we process them optimally to separate signal from noise and foregrounds? These complications are evident in all kinds of observational facilities, from multi-object spectrographs to particle detectors.

A modern multi-object spectrograph carries hundreds to thousands of fibres\footnote{Instruments like SDSS \citep{York2000_SDSS_technical, Gunn2006_SDSS_telescope}, DESI \citep{DESI2016_experiment, DESI2022_instrument}, 4MOST \citep{deJong2019_4MOST}, WEAVE \citep{Dalton2014_WEAVE, Jin2024_WEAVE}, MOONS \citep{Cirasuolo2020_MOONS}, or PFS/SuNSS on Subaru \citep{Takada2014_PFS, 2024SPIE13096E..79J}.} across a single field, and a full survey comprises thousands of visits. For each visit an observer must decide which candidates every fibre will target from a set of possibilities many orders of magnitude above the number of fibres. The classical answer treats the decision as combinatorial optimisation and solves it by hand-built machinery. The classical SDSS \emph{tiling} algorithm nudges evenly spaced pointings to smooth the sampling of a clustered sky, then allocates fibres around the collisions \citep{Blanton2003_SDSS_tiling}; the Prime Focus Spectrograph recasts the same task as a minimum-cost network flow and solves it by linear programming \citep{Fabricius2020_PFS_netflow}; 4MOST assigns fibres probabilistically \citep{Tempel2020_fibre_assignment}. Each of these solutions has been widely applied, but suboptimally and with significant drawbacks. These limitations persist despite considerable waste of observational resources and the importance of reproducible selection functions for downstream science \citep{Wang2022_spectroscopy_resourceallocation}.

\begin{figure}[t]
\centering
\includegraphics[width=0.7\textwidth]{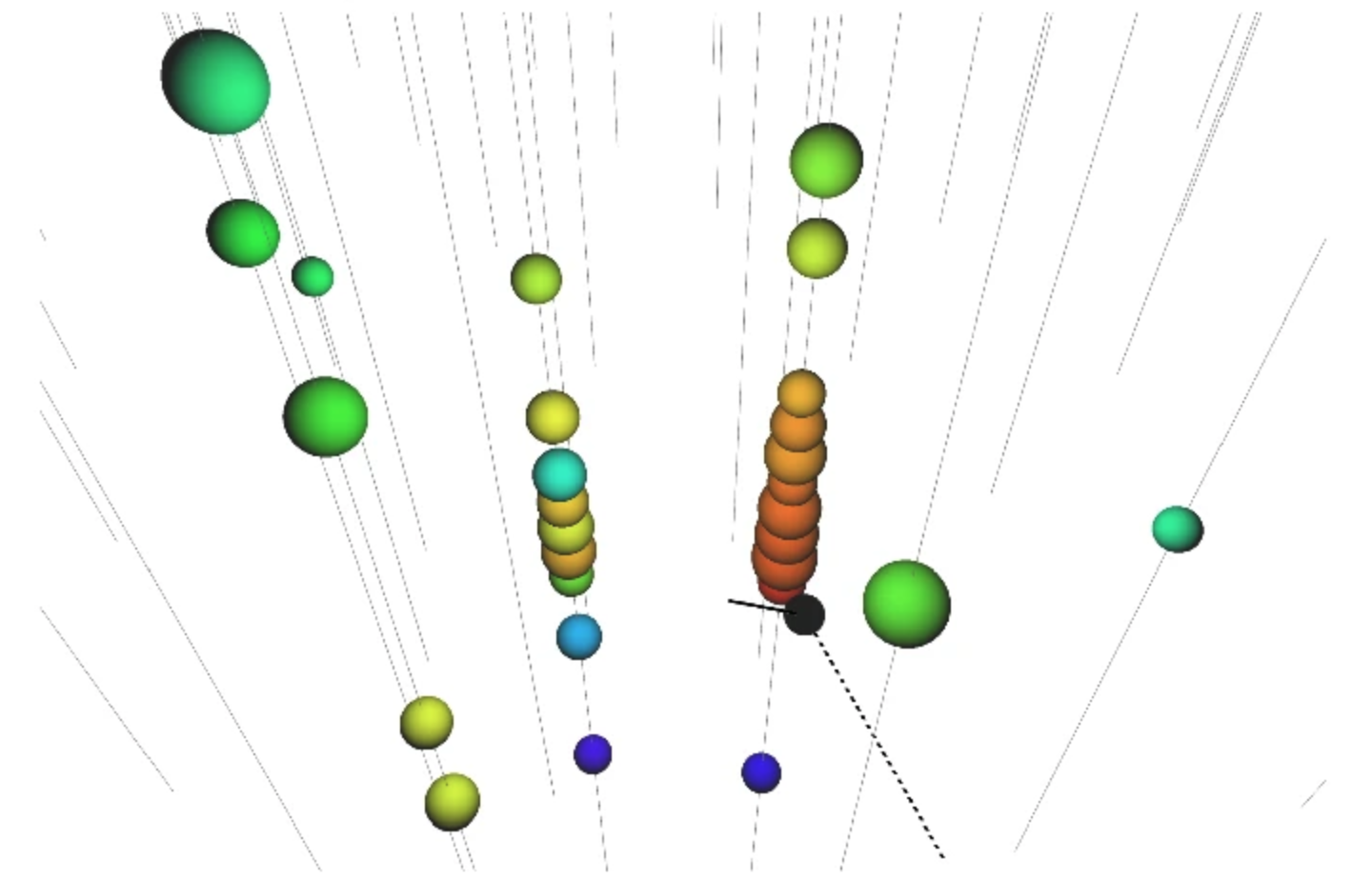}
\caption{\textbf{Example event in the IceCube detector}, the irregular sensor array that graph networks read natively. The size of spheres represents the amount of light observed in each digital optical module; larger spheres correspond to more light observed. The dotted black line represents the incoming neutrino, the black circle the cascade interaction point, and the solid black line the outgoing muon track. The size of the black round marker is proportional to the energy of the cascade at the vertex. The colour represents the timing of the hit, where the earliest hits are coloured in red and the latest hits in blue. Reproduced from \citet{Abbasi2022_DeepCore_reconstruction} with kind permission of The European Physical Journal (EPJ).}
\label{fig:icecube_graph}
\end{figure}

The hardness is inherent to the task. At its core, fibre assignment is a \emph{maximum-weight independent set} on a conflict graph, one node per candidate fibre--target pair and one edge per collision \citep{Sun2026_fiberassign_GNN}, while the tiling that feeds it belongs to the set-cover family; both problems are NP-hard, and their exact optima are beyond reach for any reasonable modern survey \citep{GareyJohnson1979_intractability}. Heuristics and relaxations therefore rule, and they are brittle: change the pipeline, add a target, or reweight the science, and much of the tuning begins again.\footnote{A further difficulty resists mathematics outright. The weights encode a social contract --- whose science to favour on a telescope that can do a program targeting Milky Way, extragalactic, or cosmological targets. No optimiser adjudicates a value judgement; at best it lays the trade-offs bare for a human to arbitrate, but the value function must be human-defined.} Graph-based machine learning has begun to shoulder the combinatorial burden. \citet{Cranmer2021_resourceallocation_unsupervised} cast resource allocation as a graph problem and optimise for cosmological constraining power, and \citet{Wang2022_spectroscopy_resourceallocation} train a graph network to allocate fibres for multi-object spectroscopy across a multi-objective survey. The fit is ideal: the conflict graph \emph{is} the problem, and a GNN amortises an optimisation the exact solver cannot afford. A network built on exactly this independent-set formulation, tested on real LAMOST fields, lands within a fraction of a percent of the optimum at survey speed \citep{Sun2026_fiberassign_GNN}, echoing the wider result that learned heuristics on graphs rival and often outperform classical combinatorial solvers \citep{Dai2017_combinatorial_GNN, Velickovic2021_algorithmic_reasoning}.

Even when the decision of what to observe is solved, we next have to deal with the structure of our detectors. Detectors rarely sit on perfect grids, and the neutrino telescope least of all --- making it ideally viewed as a graph (Figure~\ref{fig:icecube_graph}). The \emph{domain} turns inevitable at the IceCube Neutrino Observatory whose sensors hang in an irregular, two-layer, hexagonal array through the ice; a convolutional network cannot touch it without a faked grid \citep{Choma2018_IceCube_classification, Abbasi2022_IceCube_lowenergy}. GNNs reconstruct IceCube events much faster and more precisely than the classical likelihood pipeline \citep{Choma2018_IceCube_classification, Jespersen2021_icecube_bachelor, Abbasi2022_IceCube_lowenergy, Soldin2025_IceCube_GCN}, an effort the \texttt{GraphNeT} library and its benchmarks now consolidate \citep{Bukhari2023_icecube_kaggle, Sogaard2023_graphnet_neutrino, Orsoe2025_graphnet2_library, Orsoe2026_nubench_benchmark} and carry to IceCube-Gen2 and IceAct with combined transformer--GNN architectures \citep{Paul2025_IceAct_spectrum, VaraCarbonell2025_IceCubeGen2_ML}. Other neutrino telescopes, such as KM3NeT in the Mediterranean Sea, have also built signal-processing pipelines around GNNs \citep{Reck2021_GNN_MediterraneanNeutrinoObservatory}. Cosmic-ray observatories follow the same principles as neutrino observatories and can therefore adapt the same graph-based solutions for signal reconstruction \citep{Ferriere2025_GRAND_cosmicray, Rodriguez2025_Auger_photon, Macias2026_radio_instrumentation_SBI_GNN}. Laser-interferometric calibration also yields to graph-based simulation-trained networks \citep{Kannan2025_interferometer_GNN}. Going up the energy scale in the electromagnetic spectrum, the GNN also improves X-ray detectors \citep{Latorre2026_GNNs_Xray_detectors}. The same holds for gamma-ray Cherenkov detectors, where the GNN once again outperforms all other methods \citep{Glombitza2025_GNN_GammaRayCherenkov}. Wherever the instrument forsakes the Euclidean grid, the graph solves the problem.

\section{Best Practices}
\label{sec:bestpractices}

Successfully applying machine learning is not an exact science: it is closer to engineering or art. Hard-earned principles heuristically guide the development of GNNs, but practice and experimentation ultimately decide what truly works. This section gathers some of these hard-won guidelines: what sequence of layers, when to leave the graph on the shelf, how to bake in what we already know, and how to read the GNN literature well. Treat these as starting points, not commandments.

\subsection{Designing the model}

The reader who wishes to design a GNN for a given problem faces three decisions before any training run: which architecture to choose, how to embed in it what we already know through inductive biases, and whether a GNN is the right tool at all.

\subsubsection{Choosing an architecture}
\label{subsec:bp-choose}

For the first attempt, the message passing layer or the \texttt{SAGEConv} layer\footnote{Although not necessarily with the mean aggregation function.} almost always makes a sound start. The attention (GAT) layer tends to train unstably, and plain graph convolution (the \texttt{GCN} layer) often falls short on expressivity.  Although neither layer is useless, a hard time optimising the GAT or the GCN thus often says little about whether a GNN suits the task at hand, and more about which layer to use. Whether to start with message passing or \texttt{SAGEConv} depends primarily on one thing --- whether messages must be modulated and changed based on edge or receiving-node features. As an example, \texttt{SAGEConv} works well with the merger trees presented by \citet{Jespersen2022_mangrove}, since each edge encodes a uniform time step, and thus does not modulate any learned function, and a message built from the sending halo alone proved both sufficient and simpler to optimise. In short, \texttt{SAGEConv} is the right choice when messages can accurately be based only on the sending-node features. The computational efficiency of \texttt{SAGEConv} (see Figure~\ref{fig:flavours}) also makes it a good initial choice. However, as follows from Figure~\ref{fig:flavours}, the limited expressivity of \texttt{SAGEConv} (a convolutional-style layer) makes it suboptimal for many tasks. Message passing is the most expressive of the three and so, when compute and data allow, a reasonable starting point for most tasks. Although it is often a safe choice, the high expressivity makes it slow and often difficult to optimise, so reserve it for when messages must depend on edge or receiving-node features, and let \texttt{SAGEConv} serve when they need not. As an example, for the neighbourhood graphs used for dynamics, galaxy--halo connection, or halo property inference, we cannot avoid the use of a more expressive layer. A GAT should technically also work for the majority of these tasks, but we have found it to be unstable for the tasks presented here, saving little compute relative to the full message passing layer while giving up much of its expressivity. Regardless of which layer is chosen, it is wise to start with a depth of only two or three of these layers.

Since each repeated layer means one repetition of the propagation rule, the depth sets how many neighbourhood jumps information may travel. Letting this be as large as possible seems enticing, but it carries risk; the deep GNN often struggles to optimise. We will return to this in \S\ref{subsec:bp-when-not}.

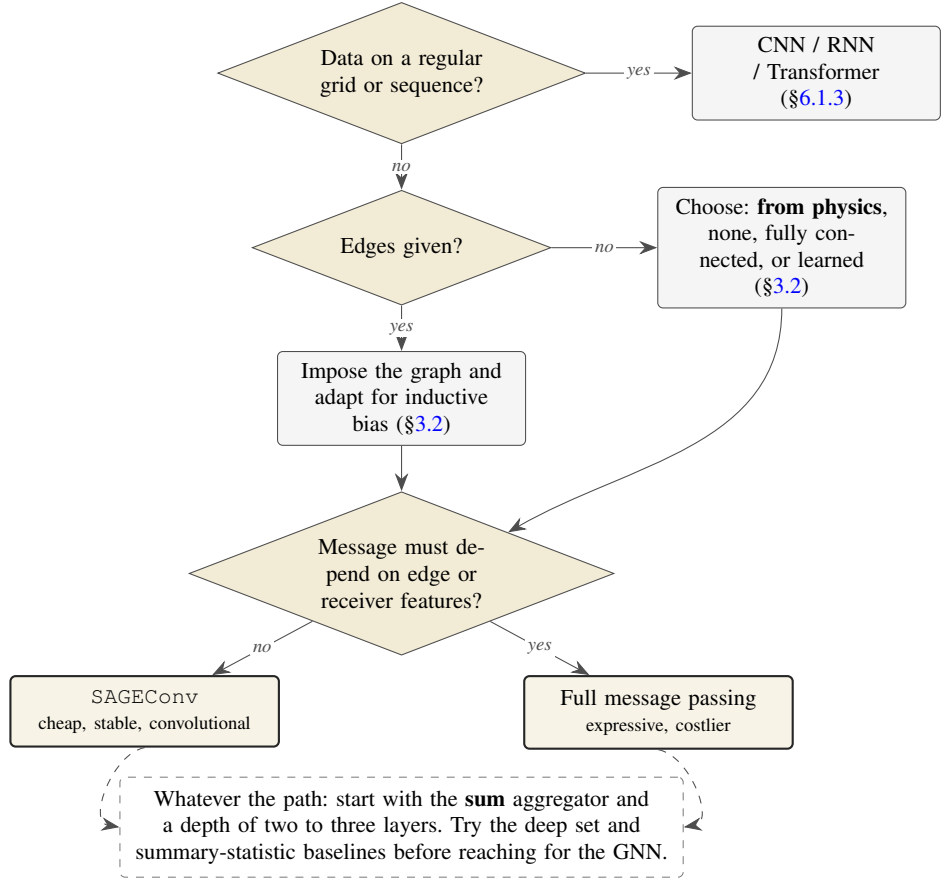
\begin{figure}
\centering
\begin{tikzpicture}[
  font=\small,
  node distance=6mm and 8mm,
  question/.style={diamond, aspect=2.6, draw=black!70, fill=rowhl, align=center, inner sep=1pt, text width=30mm},
  answer/.style={rectangle, rounded corners=2pt, draw=black!70, fill=black!4, align=center, text width=30mm, inner sep=4pt},
  final/.style={rectangle, rounded corners=2pt, draw=black!85, thick, fill=rowhl!60, align=center, text width=32mm, inner sep=5pt},
  advice/.style={rectangle, rounded corners=2pt, draw=black!50, dashed, fill=white, align=center, text width=71mm, inner sep=5pt},
  arrow/.style={-{Stealth[length=2.4mm]}, black!70},
  lab/.style={font=\scriptsize\itshape, fill=white, inner sep=1pt},
]
\node[question] (grid) {Data on a regular grid or sequence?};
\node[answer, right=14mm of grid] (cnn) {CNN / RNN / Transformer\\ (\S\ref{subsec:bp-when-not})};
\node[question, below=of grid] (edges) {Edges given?};
\node[answer, right=14mm of edges] (four) {Choose: \textbf{from physics}, none, fully connected, or learned\\  (\S\ref{subsec:no-adjacency})};
\node[answer, below=of edges] (impose) {Impose the graph and\\ adapt for inductive \\ bias (\S\ref{subsec:no-adjacency})};
\node[question, below=of impose] (msg) {Message must depend on edge or receiver features?};
\node[final, below left=8mm and 2mm of msg] (sage) {\texttt{SAGEConv}\\ \scriptsize cheap, stable, convolutional};
\node[final, below right=8mm and 2mm of msg] (mp) {Full message passing\\ \scriptsize expressive, costlier};
\node[advice, below=16mm of msg] (start) {Whatever the path: start with the \textbf{sum} aggregator and a depth of two to three layers. Try the deep set and summary-statistic baselines before reaching for the GNN.};

\draw[arrow] (grid) -- node[lab] {yes} (cnn);
\draw[arrow] (grid) -- node[lab] {no} (edges);
\draw[arrow] (edges) -- node[lab] {no} (four);
\draw[arrow] (edges) -- node[lab] {yes} (impose);
\draw[arrow] (four.south) to[out=-90, in=30] (msg.north east);
\draw[arrow] (impose) -- (msg);
\draw[arrow] (msg) -- node[lab] {no} (sage);
\draw[arrow] (msg) -- node[lab] {yes} (mp);
\draw[arrow, black!70, dashed] (sage.south) to[out=-170, in=170] (start.west);
\draw[arrow, black!70, dashed] (mp.south) to[out=-10, in=10] (start.east);
\end{tikzpicture}
\caption{\textbf{A simplified set of choices to go from data to architecture}: the decision tree summary of \S\ref{subsec:bp-choose}. Regularity sends the problem to the grid-native architectures. Irregular data pass through the graph-construction choice of \S\ref{subsec:no-adjacency}, then the layer choice splits on one central question: must the message depend on the edge or the receiver? Whatever the path, start with the sum aggregator, two to three layers, and a baseline that must be beaten. }
\label{fig:flowchart}
\end{figure}

For the neighbourhood aggregator function in the message passing layer, it is often wise to start with a sum function. A mean function is simply a normalised (and thus strictly less informative) version of the sum and is rarely the right inductive bias for physics. The same judgement extends to the canonical GCN, which normalises each aggregated message by the degrees of the sending and receiving nodes \citep{Kipf2017_GCN_semisupervised}. On a social graph the normalisation can make sense: degree measures popularity, and unless we want a few popular hubs to dominate, we may choose to normalise it out. On the proximity graph (like the cosmic web) degree measures local density, often a large part of the very signal the linking length was chosen to record. To normalise or not is therefore itself an inductive bias. When the normalisation carries physics, keep the sum. When a normalisation varies for reasons of sampling rather than physics, such as survey depth or subsampling, the mean restores comparability across graphs of very different size. The argument runs through graph size too: if we want to train a GNN on some specific average number of neighbours and deploy in a context with many more, the mean pooling can normalise the changing neighbourhood and graph sizes away. The maximum function is often a good second choice after the sum. If neither the sum nor the max works well, we may always choose to apply a set of aggregation functions (e.g. \texttt{aggregate}$~\in~\{$\texttt{sum, max, mean, min, variance}$\}$) and simply concatenate them. While this is powerful and captures the distribution of messages well, beware the increased dimensionality of the latent space. If any global pooling functions are needed for graph-level tasks, these same guidelines apply. Figure~\ref{fig:flowchart} condenses this subsection into a decision tree.

\subsubsection{Embedding inductive biases}
\label{subsec:bp-bias}

A central tenet of geometric deep learning is: \textit{``Do not learn what you already know.''}\footnote{A phrase also said countless times by Professor Peter Melchior.} Our models learn faster, more cheaply, and more robustly when we embed prior structure we trust: through the symmetries we build in, and through the graph we construct, with its resolution, locality, or causality \citep{Jespersen2022_mangrove, Chuang2024_mergertree_GNN, Wu2024_galaxyhalo_environment}. These \textit{inductive biases} --- the set of assumptions or preferences that bias our learning algorithms towards a specific solution space --- can be incredibly helpful to our learning algorithms in their search for the correct solution. This is especially important in the ``data-poor'' regime where the full space of possible solutions is too large to be efficiently constrained by the available data. None of this forbids a model without inductive biases from finding the same solution; it merely spares the model the trouble and helps with optimisation and data efficiency. As discussed in \S\ref{sec:geometry}, every algorithm carries an implicit inductive bias, but GNNs afford us increased opportunities for specifying the inductive biases ourselves.

Two concrete examples make this point. To feed the merger tree only descendant-ward edges encodes causality through a directed graph: no message may flow from a halo to its own past \citep{Jespersen2022_mangrove}.\footnote{Not enforcing causality through either the graph structure or through receiver-independent messages does not make it impossible for the GNN to learn, it simply makes it harder to find the best solution, as was found during the optimisation of the \texttt{Mangrove} architecture by \citet{Jespersen2022_mangrove}.} To hand the network only relative separations and velocities $|\mathbf{x}_i - \mathbf{x}_j|$ in place of their absolute values encodes a notion of isotropy, and that single choice carries much of the robustness of the field-level inference of \citet{deSanti2023_fieldlevel_LFI}. The choices that shrink the graph carry bias too: to cut a merger tree at a minimum progenitor mass, as in \citet{Jespersen2022_mangrove} and \citet{Chuang2024_mergertree_GNN}, encodes which histories matter. We note once again that these inductive biases are not necessary to implement in the limit of infinite data and compute. But until that is the case, we should do our best to help our algorithms learn.

With great power comes great responsibility. The prior that we impose through our choice of data processing and model architecture is fundamental for our ability to find a successful solution to any given learning problem. Even the graph structure itself stands so strong a prior that a wrong one misleads \citep{BechlerSpeicher2024_graph_overuse}. The same happens when we strictly encode the wrong symmetries in our architecture \citep{Finzi2021_RPP_softequivariance}. Strictly enforcing an inductive bias can be detrimental. When a symmetry holds only approximately, as for a survey with a preferred line of sight, redshift-space distortions, or a selection function (effects \hyperref[app:sym:deformation]{Appendix~1} formalises), a soft constraint can beat a hard one, at the cost of softer guarantees. At times, to start the optimisation from a soft inductive bias and then iteratively increase the strength of the penalty of violation can also help the model converge to the correct solution \citep{wilson2025deeplearningmysteriousdifferent}.

\subsubsection{When not to use a GNN}
\label{subsec:bp-when-not}
While a GNN can suit almost any kind of task, other methods exist for data that already carry more structure. The regular, evenly sampled one-dimensional sequence, the image on a grid, and the very large dataset of text each find a more efficient and appropriate home elsewhere. Note that \emph{regularity}, not low dimension, often makes the call: the irregular sequence of varying length sits naturally on a graph \citep{Iqbal2025_cluster_xray_GNN}, although it may also be captured with a transformer.

These considerations link to the general geometric blueprint of \S\ref{sec:geometry}. If the data have a well-specified structure, like local signals on a regular Euclidean grid, then one is better off with a CNN, and for extremely long sequences with no prior or imposable structure, the efficiency of a transformer will be crucial. The free structure and the many advantages of the GNN cost dearly: the price of its generality, of the few inherent assumptions it makes about domain or signal, runs directly to a much heavier compute bill than an inherently structure-exploiting model would pay and it is much less likely to converge to a good solution. This lack of computational efficiency is currently rarely the limiting factor for astrophysical applications, but it is a factor worth keeping in mind. However, converging to a good solution is as important in astrophysics as anywhere else. Picking the right model for the geometry of our data and signals is always worth our time and effort.

\begin{figure}[b]
\centering
\includegraphics[width=\textwidth]{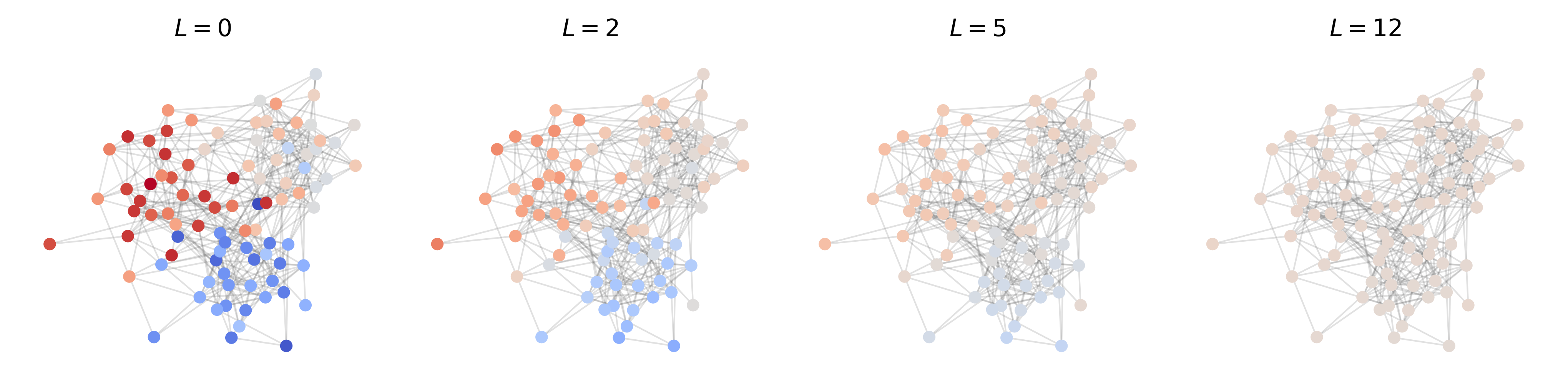}
\caption{\textbf{Oversmoothing as thermalisation}. Node features on a sample community graph, coloured by value, after zero, two, five, and twelve rounds of neighbourhood averaging: the communities blur, then vanish, as the features diffuse towards a common temperature.}
\label{fig:oversmoothing}
\end{figure}

An additional related caution follows. GNNs are constrained to rather shallow architectures, due to both the aforementioned computational inefficiency and \emph{oversmoothing} \citep{Oono2020_OverSmoothing}. Each round of message passing takes a step of diffusion, so the very deep GNN ``thermalises'': run too many rounds and every node relaxes to the same temperature, individuality boiled away into indistinguishable features (Figure~\ref{fig:oversmoothing}). Residual connections may help, and several recent works address normalisation methods to circumvent oversmoothing \citep{ruschbronstein2023surveyoversmoothing, arnaizrodriguez2026oversmoothing}. The same diffusion picture admits an opposite and rarer failure, \emph{over-squashing}: when long-range information must cross a bottleneck of few edges, too much of it is crushed into fixed-size vectors on the way through \citep{Alon2021_oversquashing_bottleneck, DiGiovanni2023_OverSquashing}. The two trade off, since the connectivity that speeds mixing also hastens equilibrium \citep{arnaizrodriguez2026oversmoothing}, but oversmoothing remains the clearer and more common issue in practice. One potential escape is to raise the body order of the message rather than the depth of the stacked layers. The MACE architecture of \citet{Batatia2022_MACE_HigherOrder_Equivariant_MP} builds many-body\footnote{A normal pairwise message aggregates direct neighbour pairs one at a time, $\sum_{j}\psi(\mathbf{x}_i,\mathbf{x}_j)$. A many-body message of order $\nu+1$ depends on $\nu$ neighbours jointly, $\sum_{j_1<\dots<j_\nu}\psi(\mathbf{x}_i,\mathbf{x}_{j_1},\dots,\mathbf{x}_{j_\nu})$.} equivariant messages through symmetrised products of the pooled neighbourhood (an expansion of classical potential theory) and so gathers in two layers the context that pairwise messages assemble over many rounds of diffusion. The network stays too shallow to thermalise, and long-range information crosses fewer bottlenecks on its way in.

The graph framing may also be used without the GNN. In \S\ref{subsec:app-stellar}, we discussed the definitional point that a graph kernel is not a GNN, but the kernel approach can be much more data-efficient. This was shown directly in \citet{Ting2023_WLkernel_chemicaltagging}, where a Gaussian-process regression with a Weisfeiler--Lehman graph kernel outperforms deep sets and graph convolutional networks on a 15-dimensional abundance space, using only $O(10)$ training simulations, two orders of magnitude fewer than the GNNs needed. The connection between the message passing GNN and the Weisfeiler--Lehman kernel is explored further in \hyperref[app:sym:wl]{Appendix~1}.

\subsection{Working with the model}
\label{sec:tutorials}

The reader who wishes to explore the use of GNNs in astrophysics and cosmology will need to engage both with hands-on practice --- with an eye on pitfalls, diagnostics, and validation --- and with the evolving literature. This is what we dive into in the next three subsections.

\subsubsection{Tutorials}
\label{subsec:tutorials}

Theory and background reading may only carry us so far; the reader who wishes to experiment finds a hands-on GNN tutorial and a companion tutorial that retraces the exploration behind a published GNN paper on the connection between a galaxy and its environment, both at \url{https://github.com/astrockragh/GNN_tutorials}, permanently archived at \href{https://doi.org/10.5281/zenodo.21926993}{doi:10.5281/zenodo.21926993}. The reader can explore how different inductive biases --- embedded through the choice of layer and aggregation --- change the performance and optimisation of the GNN. Beyond these, three external resources are worth the time. The first two are the interactive ``Gentle Introduction to Graph Neural Networks'' on Distill for beginners \citep{SanchezLengeling2021_gentle_GNN} and the Stanford CS224W course with its Colab notebooks. For the slightly more advanced reader, we cannot recommend the geometric-deep-learning proto-book and lectures of \citet{Bronstein2021_geometricDL_book} strongly enough. A full understanding of this proto-book renders one comfortable with any geometric deep learning algorithm.

\subsubsection{Pitfalls, diagnostics, and validation}
\label{subsec:bp-diagnostics}

Any model that runs will return an answer; knowing whether that answer makes sense is an art in itself. Since the graph carries so strong an inductive bias, and the GNN so much flexibility, validation deserves particular care. To earn trust, run the model through a short battery of checks before believing any conclusion (negative or positive). The following checks catch most of the failures met in practice, and each costs little.

\begin{itemize}
    \item \textbf{Split the test set.} Change and isolate it. Split by simulation box or by subgraph. When one large graph is all we have, cut it into subgraphs separated by buffer regions wider than the receptive field, as in \citet{Wu2023_galaxyenvironment_GNN} and \citet{Wu2024_galaxyhalo_environment}.
    \item \textbf{Climb the baseline ladder.} Make sure that the GNN has earned its place by fitting an MLP on hand-built summaries, then a deep set on the unordered features, and finally the GNN on the full graph. Each rung isolates one ingredient: the features, the set structure, the edges. The GNN must beat the rung below it, or the graph is just for show.
    \item \textbf{Perturb the architecture.} Vary the depth first: performance should rise, plateau, and gently fall. Track the variance of the node features across layers as well: if it collapses, the network has likely thermalised (Figure~\ref{fig:oversmoothing}), and further depth can wash out the very structure the GNN seeks to learn (\S\ref{subsec:bp-when-not}). Then jointly vary the width, the aggregator, and the layer type to potentially draw conclusions about the physics we seek to learn.
    \item \textbf{Perturb the adjacency.} The linking length or neighbour count is a hyperparameter of the physics, not only of the optimiser. Sweep it and inspect the curve, not the best point, and how sensitive the conclusion is to a choice no one can fix \textit{a priori} (Figure~\ref{fig:linking_length}).
    \item \textbf{Test the claimed or implicit symmetry.} Permute the node ordering, rotate the box, rerun. If the output drifts, the claimed symmetry does not hold.
    \item \textbf{Perturb the input features.} Drop single features and retrain, or add noise to one feature at a time. The importance ranking that emerges either matches physical expectation or teaches something, and a network that leans on a feature it should not see has found a shortcut, not a law. One can also try imputing a feature vector of just 1's (effectively no features), or randomised features. Training with either of these options can tell us a lot about what our network may or may not be learning.
    \item \textbf{Calibrate the posteriors.} When the output is a posterior, run coverage tests in the spirit of simulation-based inference \citep{Thiele2026_SBI_review}. A $\chi^2$ of the posterior may suffice in the case of a Gaussian loss function.
\end{itemize}

In short, there is no single or definitive way, but there is a roadmap of heuristic validation. We must always continue to test, calibrate, and validate.

\subsubsection{Reading the astro-GNN literature}
\label{subsec:bp-reading}

There are two general marks to look for when reading the literature on the application of GNNs (and frankly, most other deep learning methods) in astronomy and cosmology. The first is the direct line from the problem to the architecture, the geometry, and the inductive bias it invokes. It should be clear why a GNN is a suitable choice for the task. The second is a demonstrable gain, in performance or interpretability, over a simpler or more commonly applied baseline. The increased complexity of the GNN needs a demonstrable payoff.\footnote{As does any other architecture.} Beware two red flags: an invariance or equivariance claimed but not held, and the graph deployed where a grid would serve better. The first most often occurs when invariances and equivariances are not kept distinct from one another, something that happens to most physicists who instead think of both as ``symmetries'' --- a true statement, but the exact kind of symmetry is important.

\section{Frontier}
\label{sec:frontier}

Although a review focuses mostly on what has been done, the synthesis of these prior works also points beyond the present state of the art. We now flag three directions we expect to matter most: the move onto real data, generative graph models, and new kinds of graphs.

\begin{figure}[t]
\centering
\includegraphics[width=\textwidth]{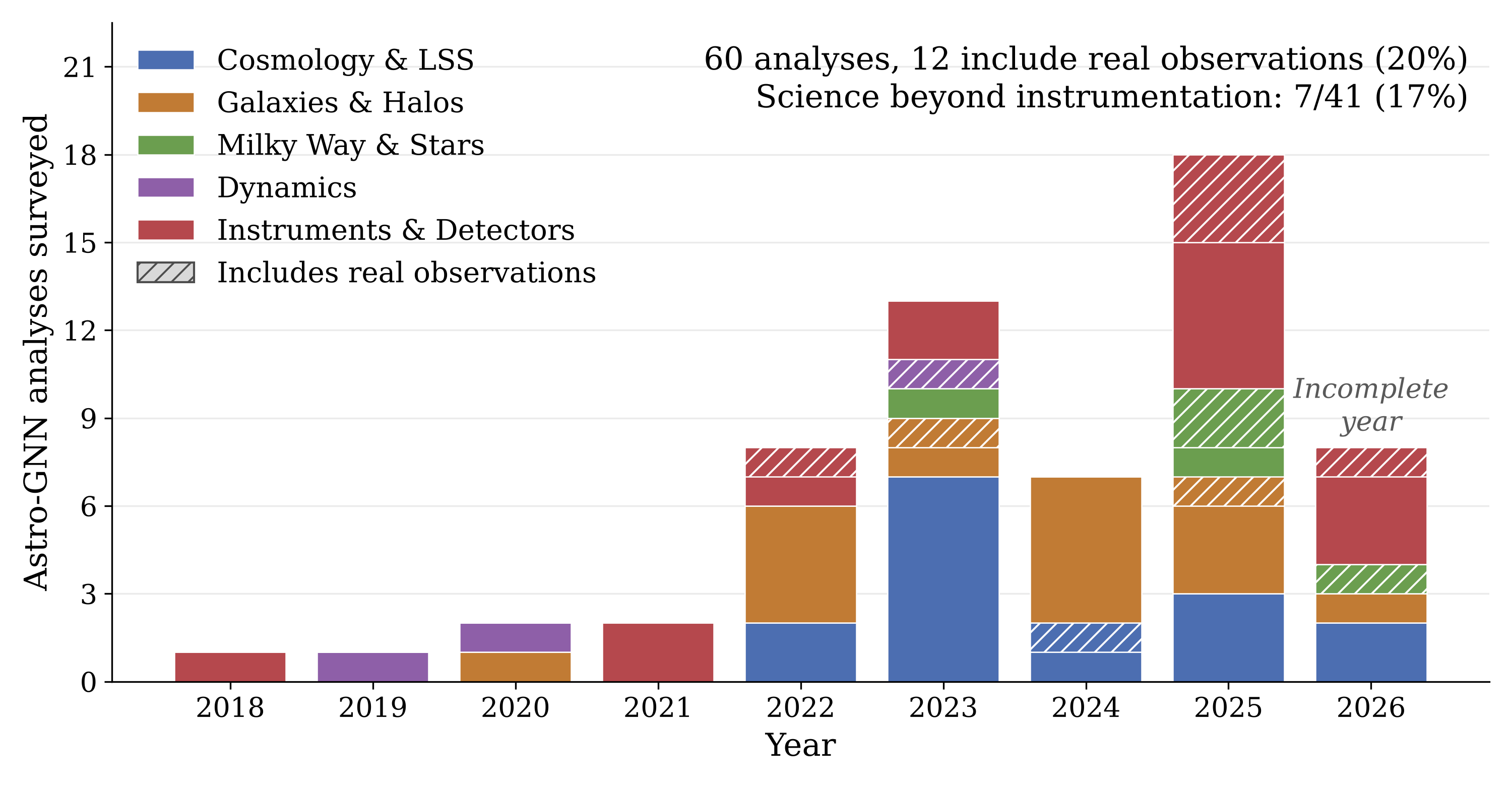}
\caption{\textbf{Astrophysical GNN analyses per year, coloured by subfield}, with hatching marking the analyses that include real observations. The simulation-to-reality gap, quantified from the bibliography of this \chapterorreview{}, is readily evident. The real-data fraction runs highest in instrumentation, detector reconstruction above all, while the scientific analyses remain overwhelmingly simulation-bound: the frontier this section argues the field must now cross. Generated from the works surveyed in \S\ref{sec:applications}.}
\label{fig:papers_per_year}
\end{figure}

\begin{itemize}
    \item \textbf{Application on real data.} The ultimate goal of natural science is experiments done using real-world data. The survey of GNN applications above reveals a clear gap (Figure~\ref{fig:papers_per_year}). Of the roughly sixty graph-network analyses we cite, one in five includes any real observations. There is a remarkable difference in data-readiness between subfields: instrumentation, detector reconstruction at IceCube and the Pierre Auger Observatory above all, includes real data in roughly one in four analyses, while the scientific analyses manage only one in six. This is the single most consequential gap in the field. It is a gap based more on general practice than on the domain-adaptive capabilities of GNNs. Any data-driven method trained on a simulation inherits the simulation's assumptions, and the move to real data demands a robustness to shifts stemming from survey masks, fibre collisions, redshift-space distortions, selection, and errors on e.g. positions, which is only now being developed. The graph framing is crucial for successfully bridging simulation and observation --- both in theory and practice. Because robust inductive biases can be included (e.g. galaxy positions enter as biased tracers rather than as pixels), the domain shift between simulation and survey typically stays smaller when using GNNs compared to similar methods, which keeps domain adaptation either tractable or unnecessary \citep{deSanti2023_fieldlevel_LFI, Roncoli2023_domainadaptation_cosmology, Lee2024_SDSS_domaingeneralization}. Porting these statistical methods onto real data is the field's most important next step. The domain gap is primed for being closed with GNNs.
    \item \textbf{Generative graph models.} A generative graph model synthesises a full graph from a lower-dimensional seed. The difficulty begins with the sheer size of the space being sampled: the number of graphs on $N$ nodes grows super-exponentially, as $2^{N(N-1)/2}$, and the model must place its mass on the vanishingly small subspace that is physically valid. But the astrophysical prize is worth that difficulty: sampling plausible merger trees or satellite environments directly from observed galaxies would be transformative. Merger-tree generation from observed properties is especially fundamental to galaxy science, since it would let us compare stellar and halo assembly directly. The astrophysical cases so far, the synthesis of merger trees and cosmic fields \citep{Tang2022_merger_normalizingflow, Jagvaral2022_intrinsicalignment_GNN, Nguyen2024_FLORAH_assembly, Nguyen2025_mergertree_generative, Nguyen2026_DREAMS_diffusion, Onutu2025_scorematching_cosmology}, already raise a second difficulty: the problems are all ill-posed, since any reasonable full graph has far more degrees of freedom than the seed vector $z$ that generates it, so the map $z \mapsto G$ is one-to-many and the model must learn a distribution $p(G \mid z)$ rather than a function \citep{Vignac2023_DiGress}. Although this structure makes the problem extremely hard, it also has a useful consequence: a forward generator $p(G\mid z)$ defines, through Bayes, its inverse $p(z\mid G) \propto p(G\mid z)\,p(z)$. Consequently, the existence of a generative model implies that we can also perform regression to recover seed parameters that gave rise to the graph. Momentum for these methods comes increasingly from outside astrophysics: graph and structure generation already designs molecules and proteins \citep{Hoogeboom2022_EDM_molecules, Watson2023_RFdiffusion}, and reward-driven samplers over graphs such as \texttt{GFlowNets} can produce diverse, high-value candidates for scientific discovery \citep{Bengio2021_GFlowNet, Jain2022_GFlowNet_biological, Bengio2023_GFlowNet_foundations}. We expect this tooling to be developed further, carried in part by heavy private investment and the economics of drug discovery, and astrophysics stands to inherit these advances.
    \item \textbf{New kinds of graphs.} The nature of the Universe is dynamic and multi-scale, meaning that the graphs themselves should ideally be able to (a) change with time and (b) capture long- and short-range interactions at once. Some solutions to capture these necessities are already in place. The \emph{topological} GNN combines graphs at different scales, capturing long- and short-range coupling without inflating the edge count to billions \citep{Lee2025_topological_cosmology}; the cosmic web is the natural target, where a galaxy's fate is spun both from its nearest neighbours and from the larger scale structures around it. The \emph{temporal} GNN, defined on a graph that rewires, grows, or shrinks with time \citep{Rossi2020_temporal_GNN}, fits time-ordered data such as the on-the-fly merger tree, which sprouts a node at every halo merger, or potentially the transient sky, whose alert graph rewires nightly as sources brighten and fade in surveys like the Rubin Observatory's LSST \citep{Ivezic2019_LSST_design}.
\end{itemize}

Unsupervised learning is surprisingly absent from the surveyed papers, despite clear successes elsewhere in astronomy on diverse tasks like clustering of gamma-ray bursts \citep{Jespersen2020_GRB_lightcurve_clustering}, outlier detection in SDSS spectra \citep{Baron2017_anomaly_SDSSspectra}, and pretraining representations that transfer to downstream tasks \citep{Hayat2021_selfsupervised_astroimages, Melchior2023_spender_autoencoder, Parker2024_AstroCLIP_crossmodal, Jespersen2026_optical_infrared}. The machinery should transfer to graphs, from graph autoencoders \citep{Kipf2016_VGAE_autoencoder} to contrastive objectives over neighbourhoods \citep{Velickovic2018_DGI_infomax}, yet no surveyed analysis is strictly unsupervised.

The path forward runs on two tracks at once. The methods already honed on simulations should now be carried onto real data, with GNN-enabled simulation-based inference as the natural vehicle \citep{Thiele2026_SBI_review}. At the same time, the methodology itself must keep advancing, through the generative models and new kinds of graphs sketched above. We expect the coming decade to move GNNs from a promising method tested mostly on simulations to a standard tool applied routinely to observations. We look forward to making this a reality.

\section{Summary and Outlook}
\label{sec:summary}

Physics advances through geometry, and the algorithms we build for it ought to do the same. We set out to review the principles and their application: fit the structure the data hold, respect the symmetries we trust, and learn only the remainder. The general geometric blueprint presented here makes these principles concrete, and the GNN stands as its most general instrument.

This \chapterorreview{} is based on three threads. First, geometry: weigh the native structure of both the domain and the signal before reaching for an architecture. Second, the graph: the relational substrate whose adjacency we define from our pre-existing physical knowledge or learn from the ground up. Third, the GNN: message passing assembled from the handful of layers and aggregators the reader can now read off and adapt at will.

The strength of geometric deep learning with GNNs lies not in their physically intuitive implementation, their embedding of inductive biases, or their adaptability alone, but in the three taken together. A GNN bakes in the symmetries and the structure we trust while fitting what we do not yet understand, and it speaks the native language of physics: centring the object and the relation between objects, encoded as points and edges, each with attributes, exchanging interaction messages to update their state vectors. This aligns with the thesis that it is impossible to fully know the parts without knowing the whole, just as it is impossible to fully know the whole without knowing the individual parts \citep{pascal_pensees_1951}.

Besides the general principles we have introduced, the best practices (\S\ref{sec:bestpractices}) and practical guidelines (\S\ref{sec:tutorials}) presented here make GNNs accessible to anyone wishing to start applying them to astrophysical data. The applications surveyed in \S\ref{sec:applications} serve as a starting point for inspiring new applications at the frontier of graph-based learning in astrophysics.

The frontier runs wide and largely open. GNNs have already been demonstrated as powerful and well-suited tools for astrophysics, but a few clear directions remain. Most importantly, the field is now ready to apply GNNs broadly to real data and to improve the scientific return on our hard-earned observational data. This is especially important given that out of the almost sixty papers surveyed here, only a small fraction actually apply GNNs to real data. The tools stand ready, and the coming years will show how much of their promise survives contact with observations. At the same time, we must continue to develop the techniques, with generative models of cosmic structure and temporal or topological graphs. Nature organises itself in objects and relations. The GNN is built to learn from both.

\begin{acknowledgement}

CKJ thanks the editors of the \textit{Machine Learning Techniques for Astrophysics and Cosmology} series, Cosimo Bambi, Vinay Kashyap, Swarnim Shashank, and Naoki Yoshida, for the invitation to write this \chapterorreview{}. CKJ wishes to thank Marielle Côté-Gendreau for ever-present help with formulating the text and arguments presented here. CKJ also deeply thanks John F. Wu and Joshua S. Speagle for valuable feedback on this \chapterorreview{}. CKJ furthermore thanks the organisers and participants of the Kavli Institute for Theoretical Physics ``Building a Physical Understanding of Galaxy Evolution with Data-driven Astronomy'' program, as well as the organisers and participants of the Centre de recherche en astrophysique du Québec (CRAQ) Summer School on Scientific Machine Learning Applied to Astrophysical Data, both of which enabled him to develop the ability to efficiently communicate the guiding principles behind Graph Neural Networks and Geometric Deep Learning. CKJ is also grateful for many useful and illuminating conversations on GNNs with Peter Melchior, Miles Cranmer, David Spergel, Shirley Ho, Paco Villaescusa-Navarro, T. Lucas Makinen, Chen-Yu Chuang, Adi Varshney, Vasily Belokurov, Tri Nguyen, Kwok Sun Tang, Natalí de Santi, Jordan Krywonos, Matthew Johnson, Yashar Hezaveh, Laurence Perreault Levasseur, and many others.

Copy-editing and writing code for constructing figures for this work were aided by Claude-Opus-5 and Fable-5. The literature search, original writing, structure, ideas, and validation of this work are entirely due to the author.
\end{acknowledgement}

\ifarxiv\printreferences\fi

\clearpage

\section*{Appendix 1: Symmetries, Representations, and Invariance}
\addcontentsline{toc}{section}{Appendix 1: Symmetries, Representations, and Invariance}
\label{app:symmetries}

This appendix collects the geometric vocabulary the main text leans on. We follow the framing of \citet{Bronstein2021_geometricDL_book}, but keep the mathematics light and the physics close; readers wanting the full group-theory machinery will find a physicist-friendly treatment in \citet{Zee2016_grouptheory_nutshell}.

\subsection*{Symmetries, representations, and invariance}
\label{app:sym:basics}

Physicists meet symmetries before they meet the word. We are often taught that ``does not care'' is a symmetry, but that only describes invariances. We tend to forget equivariances when talking about symmetries. Symmetries compose: doing one and then another is again a symmetry, every symmetry can be undone, and ``do nothing'' is a symmetry.\footnote{Called the identity.} Those three facts are exactly the axioms of a \emph{group}, the natural home for a system's symmetries. We keep the definition formal, since everything else hangs off it:

\begin{definition}[Group]
A \emph{group} is a set $G$ with a composition rule
$\circ : G\times G \to G$ (written $g\circ h = gh$) obeying:
\begin{itemize}
  \item \textbf{Associativity:} $(gh)k = g(hk)$ for all $g,h,k\in G$;
  \item \textbf{Identity:} a unique $e\in G$ with $eg = ge = g$ for all $g$;
  \item \textbf{Inverse:} each $g$ has a unique $g^{-1}$ with $gg^{-1}=g^{-1}g=e$;
  \item \textbf{Closure:} $gh\in G$ for all $g,h\in G$.
\end{itemize}
\end{definition}

Commutativity is \emph{not} required: in general $gh \neq hg$ (rotating then translating differs from translating then rotating). Groups for which $gh=hg$ always holds are called \emph{Abelian}. Large groups are often built from a handful of \emph{generators} --- e.g.\ the symmetries of an equilateral triangle are generated by one rotation and one reflection, and that same six-element group is also the group of permutations of three labelled corners. This little coincidence is the seed of a recurring theme: the \emph{same} group can act on very different objects.

\runinhead{Group actions and representations.} We rarely care about a group in the abstract; we care about how it \emph{acts} on our data. If the group acts on the underlying domain $\Omega$ (rotating points in space, relabelling nodes of a graph), it automatically acts on signals defined over that domain (rotating an image, permuting node features). The actions that matter for us are \emph{linear}, and a linear action is precisely a \emph{representation}: a rule $\rho$ that assigns to each group element $g$ an invertible matrix $\rho(g)$ such that $\rho(gh)=\rho(g)\rho(h)$ --- composing transformations becomes multiplying matrices. The physicist's reflex is the right one here: this is exactly how a scalar, a vector, or a higher tensor ``transforms under rotation'' (the rotation matrices, or for higher spins the Wigner $D$-matrices, see \citet{Zee2016_grouptheory_nutshell}), and the dimension of $\rho$ is just the size of the feature it acts on, not of the group itself.\footnote{\citet{Zee2016_grouptheory_nutshell} develops representations from this physical standpoint.}

\runinhead{Invariance vs.\ equivariance.} Imposing a symmetry on a learned function $f$ is a powerful inductive bias: it shrinks the space of candidate functions to those that already respect the symmetry, so less has to be learned from data. Two cases recur throughout the \chapterorreview{}:

\begin{definition}[Invariance and equivariance]
A function $f$ is \emph{$G$-invariant} if $f(\rho(g)x) = f(x)$ for all $g\in G$ --- the output ignores the transformation. It is \emph{$G$-equivariant} if $f(\rho(g)x) = \rho(g)\,f(x)$ --- the output transforms in lockstep with the input.
\end{definition}

The physical reading is immediate. A molecule's total energy is rotation- and permutation-\emph{invariant} (rotate the molecule, relabel its atoms --- the energy is unmoved); the force on each atom is rotation-\emph{equivariant} (rotate the molecule and every force vector rotates with it) and permutation-\emph{equivariant} (relabel the atoms and the list of forces is relabelled the same way). The same split organises graph tasks: a graph-level prediction should be permutation-invariant, while node- and edge-level predictions should be permutation-equivariant. Convolutional layers are the canonical equivariant building block --- shifting the input shifts the output feature map by the same amount --- and a final global pooling turns equivariance into invariance. That ``stack equivariant layers, then pool'' pattern is the geometric-deep-learning blueprint discussed in the main text.

\subsection*{Deformation stability: when symmetries are only approximate}
\label{app:sym:deformation}

Exact symmetries are an idealisation. The real world is noisier: a galaxy image may be mildly warped by lensing or seeing yet still show the same galaxy. What we really want is not exact invariance to a clean group, but \emph{approximate} invariance to transformations that are \emph{close} to that group. The fix is to measure how far a transformation $\tau$ sits from the symmetry group $G$ with a cost $c(\tau)$ (zero on $G$ itself) and to ask only for \emph{deformation stability},
\begin{equation}
\| f(\rho(\tau)x) - f(x)\| \;\le\; C\,c(\tau)\,\|x\|,
\label{eq:def-stability}
\end{equation}
i.e.\ small deformations may change the output, but only a little. (Small deformations cannot simply be added to the group, because composing many small ones makes a large one --- they do not close.) This is the formal version of a demand we can sometimes make: a model should degrade gracefully, not catastrophically, when the symmetry it assumes holds only approximately, and it connects to the \emph{soft}-constraint idea presented in the main text.

\subsection*{Scale separation}
\label{app:sym:scale}

Deformation stability tames \emph{local} wobble, but on a large domain there remain far too many stable functions to learn efficiently; the escape is \emph{scale separation}, since most physical tasks organise hierarchically. The classical tool is the Fourier transform, which indexes a signal by frequency and turns convolution into multiplication, but whose descriptors are fragile: an arbitrarily small, high-frequency deformation can change them by an order-one amount, because each basis function spreads over the whole domain. \emph{Wavelets} repair this with filters localised in both space and scale, so that the error of the description grows only like the deformation itself, and their fine-to-coarse composition is precisely why useful architectures run \emph{deep} and \emph{hierarchical}, interleaving local operators with coarsening. The astrophysical resonance is hard to miss: structure formation, turbulence, and the cosmic web are all multi-scale. No equally satisfying wavelet construction exists for graphs yet. See \citet{Bronstein2021_geometricDL_book}, and the wavelet references therein (e.g.\ Mallat), for the technical development.

The graph inherits every piece of this story through the \emph{graph Laplacian} $L = D - A$, with $D$ the diagonal matrix of node degrees. $L$ acts as the discrete second derivative on the graph: apply it to a signal, and it returns zero wherever the signal runs smooth across the edges and large values wherever the signal jumps. Its eigenvectors are the graph's Fourier modes and its eigenvalues their frequencies, which cashes out the promise of Figure~\ref{fig:graph_anatomy} that the spectrum of $A$, or of its Laplacian, reveals communities, diffusion, and scale. Spectral graph convolutions filter in exactly this eigenbasis \citep{Bruna2013_spectral_networks, Defferrard2016_chebnet}, and the GCN of \citet{Kipf2017_GCN_semisupervised} descends from them as a fast local approximation. Message passing reaches the same operations without ever computing an eigendecomposition, which is why (alongside the fact that it remains more physically intuitive) the main text keeps the spatial view.

\subsection*{The Weisfeiler--Lehman ceiling}
\label{app:sym:wl}

The computer-science literature measures the expressivity of message passing against a classical yardstick, the Weisfeiler--Lehman (WL) test. The WL test attacks graph isomorphism, the question of whether two graphs share one wiring: give every node the same initial colour, repeatedly recolour each node by the multiset of its neighbours' colours, and compare the two colour histograms. Message passing performs exactly this refinement with learned functions in place of recolourings, so no message-passing network can tell apart two graphs the WL test cannot \citep{Xu2018_GIN_expressiveness}. The ceiling matters in computer science, where benchmark graphs often arrive featureless and regular, and topology is all there is to learn. It matters far less in astrophysics, where the node arrives loaded with masses, positions, and velocities: two halos rarely share a feature vector, the colours differ from the start, and the pathological cases lose their bite. The practical residue of the theory, that the sum aggregator preserves strictly more of the neighbourhood than the mean or the maximum, is folded into the advice of \S\ref{subsec:bp-choose}. The canonical $c_{ij}$ of the GCN normalises each message by the degrees of sender and receiver, a choice \S\ref{subsec:bp-choose} also weighs critically.

\section*{Appendix 2: Why a convolution is implied by translation equivariance}
\addcontentsline{toc}{section}{Appendix 2: Why a convolution is implied by translation equivariance}
\label{app:conv-proof}

The main text asserts that to demand translation equivariance of a linear layer forces that layer to be a convolution. Here we prove it. The argument takes place on a periodic one-dimensional grid of length $N$; the mathematics are the same in any dimension.

To set the stage, write the signal as $\mathbf{x}\in\mathbb{R}^{N}$ with pixels $x[n]$, and the shift by $s$ as the operator $T_{s}$, with $(T_{s}\mathbf{x})[n]=x[n-s]$ and indices taken modulo $N$. A general linear layer is a matrix $\mathbf{W}$. Equivariance demands that to shift and then apply the layer equals to apply the layer and then shift: $\mathbf{W}T_{s}=T_{s}\mathbf{W}$ for every $s$.

First, read the demand at the level of matrix entries. The entry $(T_{s}\mathbf{W})_{mn}$ picks out $W_{m-s,\,n}$, while $(\mathbf{W}T_{s})_{mn}$ picks out $W_{m,\,n+s}$, so equivariance requires $W_{m-s,\,n}=W_{m,\,n+s}$ for all $m$, $n$, and $s$. Replace $n$ by $n-s$ and the requirement becomes $W_{m-s,\,n-s}=W_{m,\,n}$: shift both indices together and the entry does not move. The matrix stays constant along its diagonals, and its entries depend only on the difference of the indices, $W_{mn}=w[m-n]$ for some vector $\mathbf{w}$. A matrix of this form is called \emph{circulant}.

Second, read the circulant matrix as an operation. Its action is
\begin{equation}
(\mathbf{W}\mathbf{x})[m] \;=\; \sum_{n} w[m-n]\,x[n] \;=\; (\mathbf{w}\ast\mathbf{x})[m],
\end{equation}
a convolution of the signal with the kernel $\mathbf{w}$. The converse holds by the same computation run backwards: every convolution is a circulant matrix, and every circulant matrix commutes with every shift. The equivalence is complete. Weight sharing, the defining trick of the CNN, is not a clever economy of parameters but the unique answer to a symmetry demand. Note what the argument does \emph{not} force: locality. The kernel may span the whole grid, and the small compact kernel of the practical CNN is a second, separate inductive bias, scale separation rather than symmetry.

Third, diagonalise. All circulant matrices share one set of eigenvectors, the discrete plane waves $e^{2\pi i kn/N}$, which are exactly the eigenvectors of the shift itself. In that basis every convolution acts as a pointwise multiplication, which is the convolution theorem, and the kernel's eigenvalues are its Fourier coefficients. This is the door through which the graph enters: on a graph the role of the shift passes to the adjacency or the role of the plane waves to their eigenvectors, and a filter diagonal in that eigenbasis is the spectral graph convolution of \hyperref[app:sym:scale]{Appendix~1}.

\section*{Appendix 3: The surveyed analyses}
\addcontentsline{toc}{section}{Appendix 3: The surveyed analyses}
\label{app:survey}

Table~\ref{tab:survey} lists every astro-GNN analysis surveyed in this \chapterorreview{}, with the classification that generates Figure~\ref{fig:papers_per_year}, and therefore includes subfield, year, and whether the analysis includes real observations. The figure, the table, and the numbers quoted in \S\ref{sec:frontier} all derive from one source list in the figure script, so the count is reproducible.

\providecommand{\redx}{\hspace{1pt}\mbox{\color{red}\ding{55}}\hspace{1pt}}
\providecommand{\greencheck}{\hspace{0.5pt}\mbox{\color{green!70!black}\ding{51}}\hspace{0.5pt}}
\begingroup\footnotesize
\setlength{\LTleft}{\fill}\setlength{\LTright}{\fill}
\begin{longtable}{l c l c}
\caption{The astro-GNN analyses surveyed in this \chapterorreview{}, as classified for Figure~\ref{fig:papers_per_year}. \greencheck{} marks an analysis that includes real observations, \redx{} one demonstrated on simulations alone.}\label{tab:survey}\\
\hline\noalign{\smallskip}
Analysis & Year & Subfield & Real obs.\\
\noalign{\smallskip}\svhline\noalign{\smallskip}
\endfirsthead
\hline\noalign{\smallskip}
Analysis & Year & Subfield & Real obs.\\
\noalign{\smallskip}\svhline\noalign{\smallskip}
\endhead
\endfoot
\noalign{\smallskip}\hline
\endlastfoot
\citet{Makinen2022_cosmicgraph_LSS} & 2022 & Cosmology \& LSS & No (\redx)\\
\citet{VillanuevaDomingo2022_cosmicgraph_clustering} & 2022 & Cosmology \& LSS & No (\redx)\\
\citet{Adams2023_CMB_graphconv} & 2023 & Cosmology \& LSS & No (\redx)\\
\citet{deSanti2023_fieldlevel_LFI} & 2023 & Cosmology \& LSS & No (\redx)\\
\citet{Jung2023_QuijotePNG_massfunction} & 2023 & Cosmology \& LSS & No (\redx)\\
\citet{Park2023_lensing_bayesianGNN} & 2023 & Cosmology \& LSS & No (\redx)\\
\citet{Roncoli2023_domainadaptation_cosmology} & 2023 & Cosmology \& LSS & No (\redx)\\
\citet{Shao2023_omegam_universalequation} & 2023 & Cosmology \& LSS & No (\redx)\\
\citet{Yang2023_haloclustering_graph} & 2023 & Cosmology \& LSS & No (\redx)\\
\citet{Kvasiuk2024_fieldreconstruction_GNN} & 2024 & Cosmology \& LSS & No (\redx)\\
\citet{Lee2024_SDSS_domaingeneralization} & 2024 & Cosmology \& LSS & Yes (\greencheck)\\
\citet{Huang2025_CosmoBench_benchmark} & 2025 & Cosmology \& LSS & No (\redx)\\
\citet{Kololgi2025_cosmicweb_classification} & 2025 & Cosmology \& LSS & No (\redx)\\
\citet{Lee2025_topological_cosmology} & 2025 & Cosmology \& LSS & No (\redx)\\
\citet{Krywonos2026_GNNCNN_reconstructing_DM_from_galaxies} & 2026 & Cosmology \& LSS & No (\redx)\\
\citet{Thiele2026_void_graphflow} & 2026 & Cosmology \& LSS & No (\redx)\\
\citet{Roper2020_MEGA_mergergraph} & 2020 & Galaxies \& Halos & No (\redx)\\
\citet{Jagvaral2022_intrinsicalignment_GNN} & 2022 & Galaxies \& Halos & No (\redx)\\
\citet{Jespersen2022_mangrove} & 2022 & Galaxies \& Halos & No (\redx)\\
\citet{Tang2022_merger_normalizingflow} & 2022 & Galaxies \& Halos & No (\redx)\\
\citet{VillanuevaDomingo2022_halomass_GNN} & 2022 & Galaxies \& Halos & No (\redx)\\
\citet{VillanuevaDomingo2023_MW_mass} & 2023 & Galaxies \& Halos & Yes (\greencheck)\\
\citet{Wu2023_galaxyenvironment_GNN} & 2023 & Galaxies \& Halos & No (\redx)\\
\citet{Chuang2024_mergertree_GNN} & 2024 & Galaxies \& Halos & No (\redx)\\
\citet{Garuda2024_GNN_reconstructing_halomass_from_galaxies} & 2024 & Galaxies \& Halos & No (\redx)\\
\citet{Larson2024_halomass_images} & 2024 & Galaxies \& Halos & No (\redx)\\
\citet{Nguyen2024_FLORAH_assembly} & 2024 & Galaxies \& Halos & No (\redx)\\
\citet{Wu2024_galaxyhalo_environment} & 2024 & Galaxies \& Halos & No (\redx)\\
\citet{Delgado2025_cluster_triaxiality} & 2025 & Galaxies \& Halos & No (\redx)\\
\citet{Iqbal2025_cluster_xray_GNN} & 2025 & Galaxies \& Halos & Yes (\greencheck)\\
\citet{Nguyen2025_mergertree_generative} & 2025 & Galaxies \& Halos & No (\redx)\\
\citet{Onutu2025_scorematching_cosmology} & 2025 & Galaxies \& Halos & No (\redx)\\
\citet{Nguyen2026_DREAMS_diffusion} & 2026 & Galaxies \& Halos & No (\redx)\\
\citet{Nguyen2023_dwarf_densityprofile} & 2023 & Milky Way \& Stars & No (\redx)\\
\citet{Berni2025_MWhalo_substructure} & 2025 & Milky Way \& Stars & Yes (\greencheck)\\
\citet{Nguyen2025_dwarf_GraphNPE} & 2025 & Milky Way \& Stars & No (\redx)\\
\citet{Spina2025_chemicaltagging_GAT} & 2025 & Milky Way \& Stars & Yes (\greencheck)\\
\citet{Nguyen2026graphNPE_applied} & 2026 & Milky Way \& Stars & Yes (\greencheck)\\
\citet{Cranmer2019_symbolicphysics_GNN} & 2019 & Dynamics & No (\redx)\\
\citet{Cranmer2020_symbolicregression_inductivebias} & 2020 & Dynamics & No (\redx)\\
\citet{Lemos2023_orbitalmechanics_symbolic} & 2023 & Dynamics & Yes (\greencheck)\\
\citet{Choma2018_IceCube_classification} & 2018 & Instruments \& Detectors & No (\redx)\\
\citet{Cranmer2021_resourceallocation_unsupervised} & 2021 & Instruments \& Detectors & No (\redx)\\
\citet{Reck2021_GNN_MediterraneanNeutrinoObservatory} & 2021 & Instruments \& Detectors & No (\redx)\\
\citet{Abbasi2022_IceCube_lowenergy} & 2022 & Instruments \& Detectors & Yes (\greencheck)\\
\citet{Wang2022_spectroscopy_resourceallocation} & 2022 & Instruments \& Detectors & No (\redx)\\
\citet{Bukhari2023_icecube_kaggle} & 2023 & Instruments \& Detectors & No (\redx)\\
\citet{Sogaard2023_graphnet_neutrino} & 2023 & Instruments \& Detectors & No (\redx)\\
\citet{Ferriere2025_GRAND_cosmicray} & 2025 & Instruments \& Detectors & No (\redx)\\
\citet{Glombitza2025_GNN_GammaRayCherenkov} & 2025 & Instruments \& Detectors & No (\redx)\\
\citet{Kannan2025_interferometer_GNN} & 2025 & Instruments \& Detectors & No (\redx)\\
\citet{Orsoe2025_graphnet2_library} & 2025 & Instruments \& Detectors & No (\redx)\\
\citet{Paul2025_IceAct_spectrum} & 2025 & Instruments \& Detectors & Yes (\greencheck)\\
\citet{Rodriguez2025_Auger_photon} & 2025 & Instruments \& Detectors & Yes (\greencheck)\\
\citet{Soldin2025_IceCube_GCN} & 2025 & Instruments \& Detectors & Yes (\greencheck)\\
\citet{VaraCarbonell2025_IceCubeGen2_ML} & 2025 & Instruments \& Detectors & No (\redx)\\
\citet{Latorre2026_GNNs_Xray_detectors} & 2026 & Instruments \& Detectors & No (\redx)\\
\citet{Macias2026_radio_instrumentation_SBI_GNN} & 2026 & Instruments \& Detectors & No (\redx)\\
\citet{Orsoe2026_nubench_benchmark} & 2026 & Instruments \& Detectors & No (\redx)\\
\citet{Sun2026_fiberassign_GNN} & 2026 & Instruments \& Detectors & Yes (\greencheck)\\
\end{longtable}
\endgroup

\ifarxiv\else\printreferences\fi%

\end{document}